\documentclass{article}
\usepackage{graphicx}

\def\T{{\scriptscriptstyle T}}
\def\V{\mathbf}
\def\Pe{{\rm Pe\/}}
\def\lqcd{\Lambda_{\scriptscriptstyle QCD}}
\def\ppbar{\left\langle\overline\psi\psi\right\rangle}

\def\etal{{\sl et al.\/}}
\def\etc{{\sl etc.\/}}
\def\ie{{\sl i.e.\/}}
\def\eg{{\sl e.g.\/}}

\def\beqa{\begin{eqnarray*}}
\def\eeqa{\end{eqnarray*}}
\newcommand{\beq}{\begin{equation}}
\newcommand{\eeq}{\end{equation}}
\newcommand{\bef}{\begin{figure}}
\newcommand{\eef}{\end{figure}}

\title{A short introduction to heavy-ion physics}
\author{Sourendu Gupta\\Department of Theoretical Physics\\
 Tata Institute of Fundamental Research,\\
 Homi Bhabha Road, Mumbai 400005, India}
\date{{\small Lectures delivered at the Asia-Europe-Pacific School of
 High-Energy Physics,\\ Puri, India, November 2014.}}

\begin{document}

\maketitle

\begin{flushright}
TIFR/TH/15-24
\end{flushright}

\begin{abstract}
Heavy-ion collisions provide the only laboratory tests of relativistic
quantum field theory at finite temperature. Understanding these is
a necessary step in understanding the origins of our universe. These
lectures introduce the subject to experimental particle physicists, in the
hope that they will be useful to others as well. The phase diagram of QCD
is briefly touched upon. Kinematic variables which arise in the collisions
of heavy-ions beyond those in the collisions of protons or electrons are
introduced. Finally, a few of the signals studied in heavy-ion collisions,
and the kind of physics questions which they open up are discussed.
\end{abstract}

\section{Why study heavy-ion collisions}

The universe started hot and small, and cooled as it expanded.  Today vast
parts of the universe are free of particles, except for photons with
energy of about 3 Kelvin or lower. This energy scale is so far below the
mass scale of any other particle that no scattering processes occur in
this heat bath of photons. So we may consider them to be free.

This was not always so. Earlier in the history of the universe the
temperature $T$ was comparable to, or larger than, the mass scales of
many particles. As a result particle production and transmutations were
common. In those circumstances would it be correct and useful to treat
this fluid as an ideal gas? Such a gas cannot give rise to freeze out,
phase transitions or rapid crossovers, and transport.  We see the
signatures of several such phenomena today, so we know that the ideal
gas treatment would not work at all times.

In the early universe many of the component particles of the fluid were
relativistic. Since we wish to describe particle production processes
in this fluid, we are forced to use quantum field theory at finite
temperature to describe the contents of the early universe. The main
theoretical tools required to study thermal quantum field theory (TQFT)
are effective field theory (which includes hydrodynamics and transport
theory) and lattice field theory.  Perturbation theory plays a limited
but very important role, due to our detailed understanding of the
technique. In order to test the formulation of TQFT, we need to think
of experiments which can be performed easily.

Experimental tests of TQFT in the electro-weak sector turn out to be
unfeasible.  Initial states made of leptons may achieve energy densities
of the order of 1/fm$^4$. However, mean-free paths due to electro-weak
interactions are of the order 100 fm. So it is very hard to thermalize
this energy density.  Initial states of hadrons, on the other hand,
have mean-free paths of the order of 1 fm, so the initial energy may be
converted into thermal energy.  By using heavy-ions, one can increase
the initial volume significantly, and so improve the chances of producing
thermalized matter. This is why heavy-ion collisions (HICs) are used to
test TQFT.

The objects of experimental study should be as many as possible, in
order to subject TQFT to as many tests as can be conceived. The most
important phenomena are transport properties: the electrical conductivity
(important for the freezeout of photons), viscosity (responsible for
entropy production), the speed of sound, the equation of state and so
on. But perhaps the most interesting objects of experimental study are the
possible phase transitions and crossovers associated with the symmetries
of the standard model. Corresponding to every global symmetry there is
a chemical potential. So the phase diagram of the standard model has
high dimensionality and potentially many phases. Experiments which are
feasible in colliders can reach only a small fraction of the phases.

\section{Symmetries and states of QCD}

Phase diagrams display the conditions under which global symmetries are
broken or restored. Heavy-ion collisions explore the phase diagram of QCD.
The global symmetries of this theory are chiral SU$_L$($N_f$) $\times$
SU$_R$($N_f$) $\times$ U$_B$(1), where $N_f$ is the number of flavours
of light quarks, the subscripts $L$ and $R$ stand for left and right
chirality, and B for baryon number.

Chiral symmetry is explicitly broken by quark masses. QCD contains a
scale, $\lqcd$. Quarks with masses larger than $\lqcd$ are far from
the chiral limit. The strange quark mass, $m_s$, is near the scale of
$\lqcd$, and it is a detailed question whether treating it as nearly
chiral helps in understanding the phenomenology of strong interactions.
The up and down quark masses are much lighter than $\lqcd$ and it is
useful to treat them as nearly chiral.

The resulting SU$_L$(2) $\times$ SU$_R$(2) {\sl chiral symmetry
is spontaneously broken\/} down to SU(2) isospin symmetry in the
vacuum. Signals of this symmetry breaking are the fact that the QCD vacuum
contains a non-vanishing chiral condensate, $\ppbar$, and that pions
are massless.  Departures from chirality are important and treated in
chiral perturbation theory \cite{weinberg}: the most important result is
that pions get a mass proportional to the square root of the quark mass.

\bef[t]
\begin{center}
\includegraphics[scale=0.4]{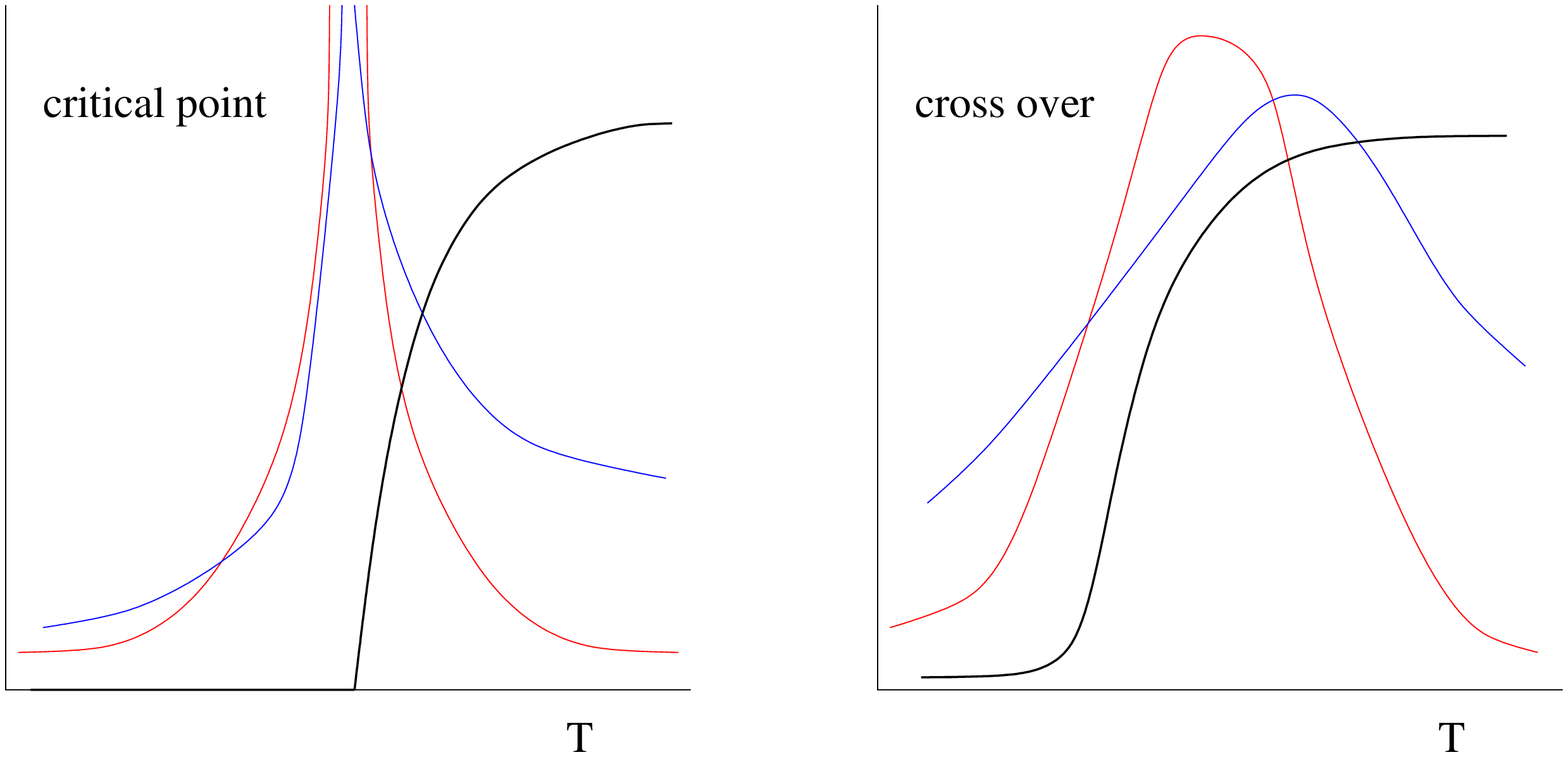}
\end{center}
\caption{At a critical point order parameters change abruptly, and
 specific heats and susceptibilities may have singularities. The location
 of the singularity is unique. At a crossover there are no singularities.
 The order parameters may have a large continuous change. It is possible
 that specific heats and susceptibilities peak as the temperature changes.
 The locations of maximum slope of the order parameter, or the peaks of
 susceptibilities, generally depend on the observable chosen.}
\label{fg.crossover}\eef

As the temperature of the vacuum is raised, keeping the baryon number
and charge densities at zero, the condensate changes to a very small
value, proportional to the quark mass.  From thermodynamic arguments,
models, and lattice QCD computations it is known that the change is
gradual (see Figure \ref{fg.crossover}). One may try to characterize
a temperature where this crossover happens, but it is a conventional
number \cite{crossover}.  The {\sl crossover temperature\/}, $T_c$,
depends upon which physical quantity is examined, but it is perfectly
well-defined after a choice is made\footnote{Another example of a
crossover is the formation of a glass by cooling of liquid silica. The
glass transition temperature depends on what measurement one makes on
the sample of the glass.} We make the choice that $T_c$ is given by the
peak of the Polyakov loop susceptibility.

The quark number for each of the $N_f$ flavours is conserved. For the
study of the phase diagram we need to keep in mind the up and down quark
numbers (or, equivalently, the baryon number and the net isospin). A
grand-canonical ensemble for QCD would then need two chemical potentials,
$\mu_u$ and $\mu_d$ (or $\mu_B$ and $\mu_I$), and the temperature $T$:
so the phase diagram is three dimensional. As a first approximation one
treats the up and down quark masses to be equal, and examines the phase
diagram in the two dimensional slice with $T$ and $\mu_B$ \cite{phased},
and independently, of that in $T$ and $\mu_I$ \cite{isospin}. There has
been little study of the more complete (and complicated) phase diagram
\cite{sgupta}.

The phase diagram in $T$ and $\mu_B$ for small $\mu_B$ and non-zero
light quark mass (see Figure \ref{fg.phased}) was first investigated in
\cite{phased}. At small $\mu_B$ the states of QCD are distinguished by
the value of the chiral condensate.  In the low $T$ state it is large,
but becomes small at $T>T_c$.  At sufficiently high $T$ the dependence
of the condensate on $\mu_B$ is computable, and shows a gradual
variation. However, various arguments lead to the expectation that at
low $T$ as one changes $\mu_B$ there is a first order phase transition
signalled by an abrupt change in the condensate. The thermodynamic Gibbs'
phase rule \cite{landau} then tells us that the phase diagram has a line
of first order transition.  As we already discussed, this cannot hit the
$\mu_B=0$ axis or rise to $T\to\infty$. So it must end somewhere. The end
point is a second order phase transition, called the QCD critical point.

The actual location of the curve of first order phase transition curve
and the critical end point can only be predicted by a non-perturbative
computation, \ie, through {\sl lattice QCD simulations\/}
\cite{lattice}. However, a computation at finite $\mu_B$ requires an
extension of known techniques because of a technical problem known as
the {\sl fermion sign problem\/}. Many such methods have been proposed,
and many are being explored \cite{lattice}. It would be fair to say that
developing such techniques is one of the most active areas in lattice
gauge theory today.

\bef[t]
\begin{center}
\includegraphics[scale=0.4]{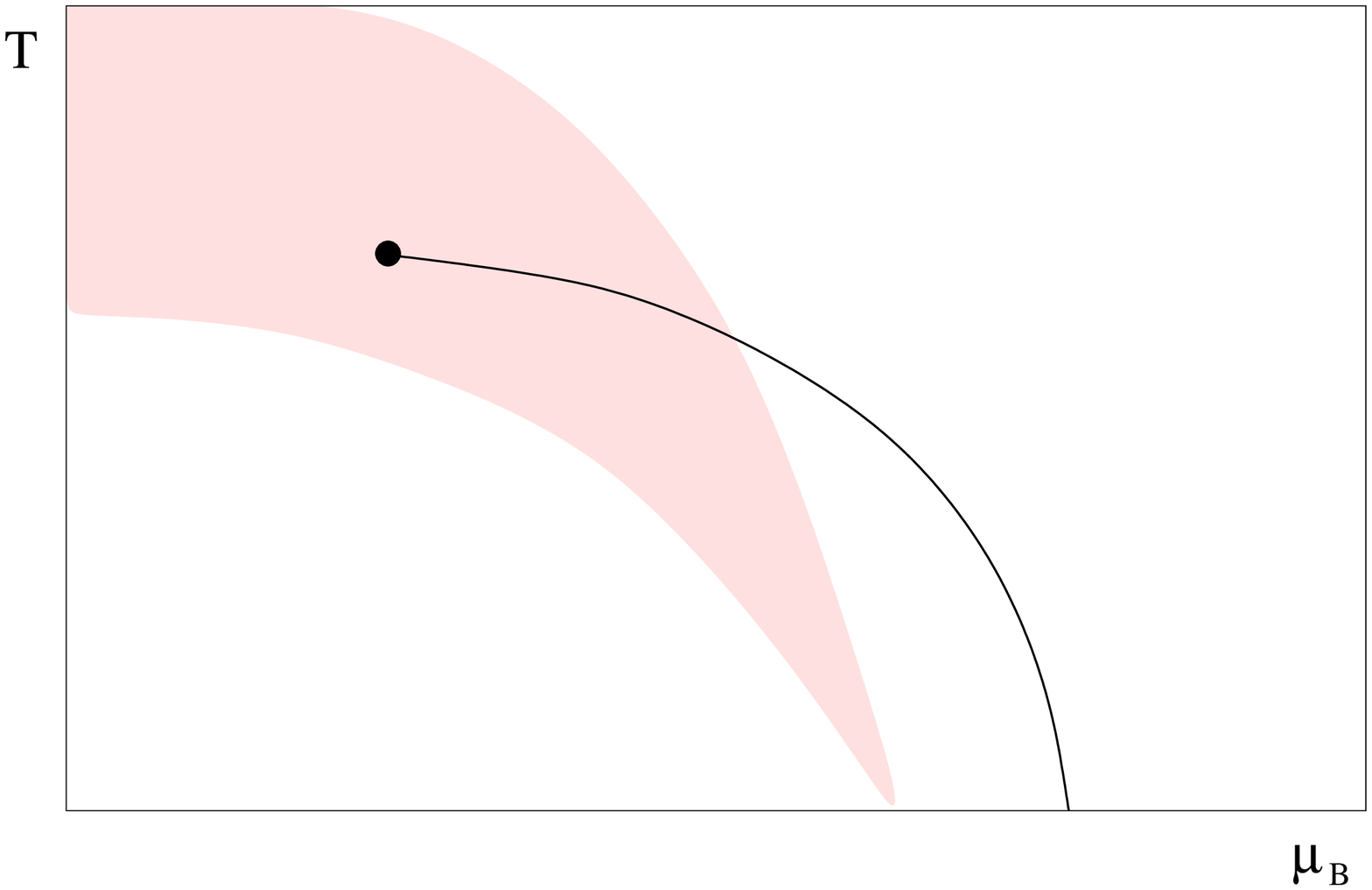}
\end{center}
\caption{The phase diagram of QCD in the $T$-$\mu_B$ plane as seen with the
 chiral condensate \cite{phased}. A line of first order phase transitions
 (black line) ends in the QCD critical point (black dot). The fireball
 produced in a heavy-ion collision lives in a track within the shaded domain.
 The lower edge of this domain is called the freezeout curve. The domain is
 traced out by tracks of the history of the fireball as the collider energy
 changes. For small energies, the domain ends at $T=0$ and a chemical
 potential corresponding to nuclear matter. As the energy increases, the
 domain moves to small $\mu_B$. The logic of the beam energy scan is that
 this domain is likely to include the QCD critical point.}
\label{fg.phased}\eef

Till now extensive computations in QCD with varying lattice cutoffs,
spatial volumes and quark masses has been possible using only one
particular method, involving the Taylor expansion of the pressure in
powers of $\mu_B$.  As a result the information available until now is
fairly limited, and one would hope that the future brings alternative
computational schemes. The current best estimate of the location of the
critical point is \cite{ilgti}
\beq
   T^E \simeq (0.94\pm0.01) T_c  \qquad{\rm and}\qquad 
   \mu_B^E \simeq (1.68\pm0.06) T^E.
\label{cep}\eeq
Methods have also been developed to compute the equation of state, the
bulk compressibility, and the speed of sound in several parts of the phase
diagram \cite{ilgti}.

Two more aspects of the phase diagram of QCD are interesting, but cannot
be described here. One is the temperature dependence of the axial anomaly.
This has been keenly investigated in recent years \cite{axial}. The other
is the phase diagram of QCD in a strong and constant external magnetic field.
This has also generated much work recently \cite{magfield}.

\section{General conditions in heavy-ion collisions}

I turn now to heavy-ion collisions, which is the experimental system that
can test the computations we discussed briefly in the previous section.
In this section I touch upon three related questions: whether thermal
matter is produced, what its flavour content is likely to be, and how
one can control the energy content of this matter.

The object of study in heavy-ion collisions is the (hopefully) thermalized
matter in the final state. In high energy colliders matter is always
formed.  In sufficiently hard pp collisions, for example at the LHC,
even soft physics contains enough energy to create W/Z bosons, not to
speak of hadrons. The mere production of large amounts of hadronic matter
is not of interest. What we need to know is whether this matter
re-interacts with sufficient strength to thermalize. In the language of
particle physics this is about final state effects.

In order to understand the time scales involved it is sufficient to run
through a simple kinetic theory argument.  Let the two-body scattering
cross section be $\sigma$. Taking the number density of particles in
the final state to be $n$, one can write the mean free path as
\begin{equation}
   \lambda \propto (n\sigma)^{-1}, 
\label{mfp}\end{equation}
If the dimensionless number $1/(\lambda\sqrt[3]{n}) = \sigma/n^{2/3} =
{\cal O}(1)$ then the mean free path is of the same order as the mean
separation between particles. In this case, final state collisions are
numerous, and matter may come into {\sl local thermal equilibrium\/}.

When $\sqrt S\simeq20$ GeV, we know that jets are rare. As a result, we
can take the final state particles to be hadrons, so that $\sigma\simeq40$
mb. In this case $n\ge 5$/fm${}^3$ may be sufficient for the final state
to thermalize. This number density cannot be reached in collisions of
protons. However, heavy-ion collisions increases $n$ by some power of
$A$, so heavy-ion collisions at this energy may thermalize. At the LHC,
$n$ is large, so thermalization is easier. Even high multiplicity
pp collisions may then thermalize. The thermalized system arising from these
collisions is the {\sl fireball\/} which is the object of study in
heavy-ion collisions.

This treatment is sufficient for building intuition, but a quantitative
analysis of thermalization is more complex. The rapid expansion of the
fireball implies that simple kinetic theory does not suffice, and the
theoretical framework becomes more complex.  Some relevant references
are collected here \cite{thermalize}.

The flavour content of the fireball is needed in many analyses. Again,
simple arguments are sufficient to gain a quick intuition about this.
The flavour quantum numbers of the incoming hadrons are essentially
contained in hard (valence) quarks. At large $\sqrt S$, the asymptotic
freedom of QCD guarantees that our intuition about Rutherford scattering
holds, and these valence quarks do not undergo large
angle scattering. As a result, the incoming quantum numbers are mostly
carried forward into the fragmentation region. 
In terms of the pseudo-rapidity
\begin{equation}
   \eta=\frac12\log\tan\theta,
\label{eta}\end{equation}
(where $\theta$ is the scattering angle) the {\sl fragmentation region\/} is
the region of large $|\eta|$, and is called so because (classically) one
finds the unscattered fragments of the initial particles here.

Although the valence partons individually contain large momenta,
there are only three of them in a baryon. Soft (sea) partons are much
more numerous.  As a result, quite a significant fraction of the energy
is carried by all the soft partons together. These generally scatter by
large angles and so stay in the {\sl central rapidity region\/} (\ie, the
region with $|\eta|\simeq1$).  If this matter approximately thermalizes,
then it makes the fireball which is the object of heavy-ion studies.
The net-baryon and flavour content is small, the energy content increases
with $\sqrt S$.  At high energies the central and fragmentation regions
are expected to be well separated, \ie, one expects few hadrons in the
intermediate region between them.

At $\sqrt S\simeq$1--10 GeV, baryon interactions cannot be analyzed in terms
of quarks. In this regime the fireball may contain baryon and other
flavour quantum numbers. The distinction between fireball (central)
and fragmentation region may be weak.

In the collision of point-like particles in quantum theory, the
observables are the number of particles (or energy-momentum) hitting the
detector at any angle $\theta$. The only control parameter is the center
of mass energy, $\sqrt S$.  In collisions of extended objects, there is
another control parameter: the impact parameter, $b$. This measures the
separation between the centers (geometrical centers, centers of energy)
of the colliding objects. However, $b$ cannot directly be measured in
an experiment.

\bef[ht]
\begin{center}
\includegraphics[scale=0.4]{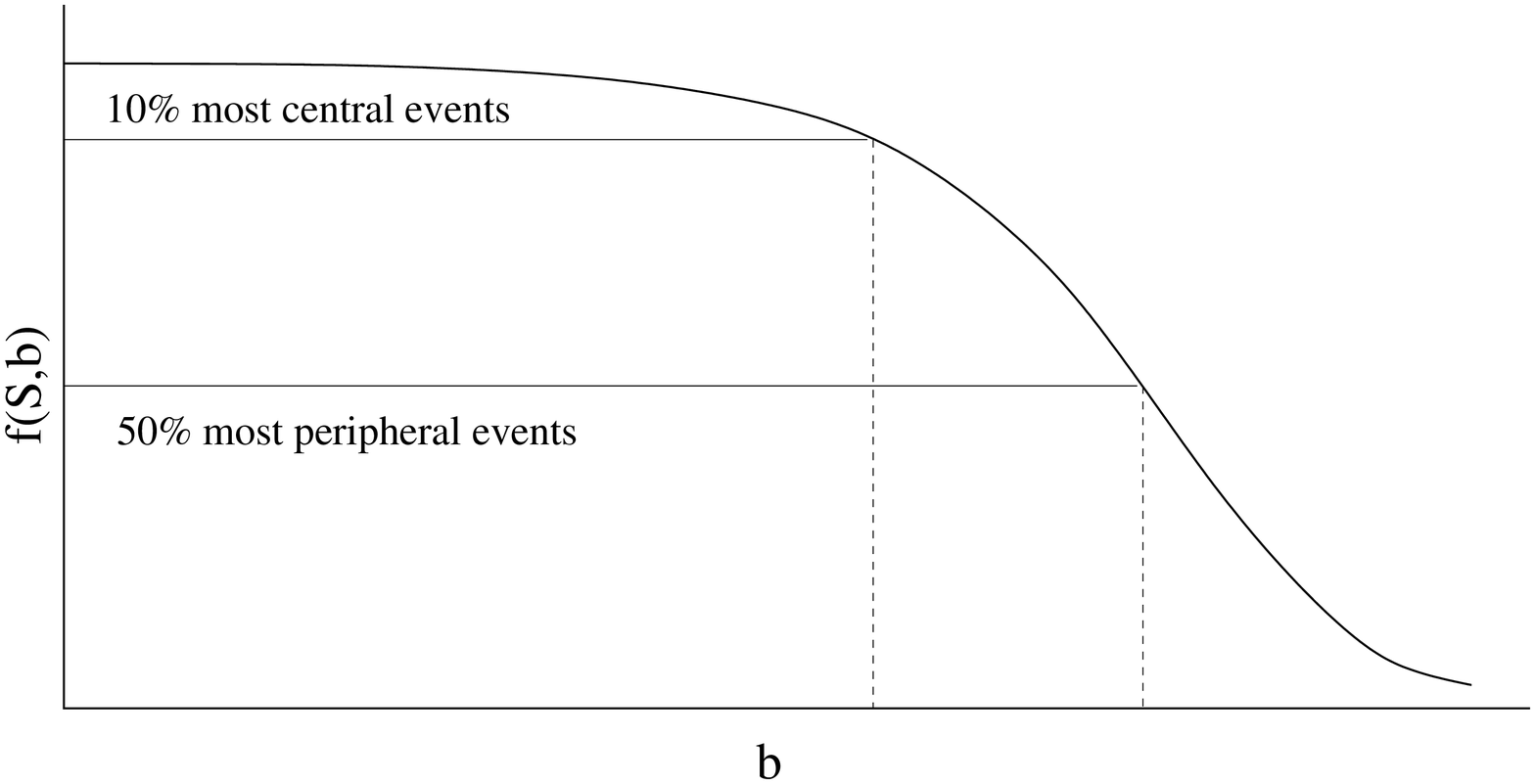}
\end{center}
\caption{The fraction of the total cross section can be used to define
 centrality classes. Different models of nuclear densities will map a
 centrality class to different impact parameter ranges.}
\label{fg.centrality}\eef

Instead we perform the following analysis. The total nucleus-nucleus cross
section depends only on the energy, so one has
\beq
   \sigma(S) = \int_0^\infty db \frac{d\sigma(S)}{db}.
\label{totcs}\eeq
Since cross sections are non-negative, the fractional cross section
\[ f(S,b) = \frac1{\sigma(S)}
          \int_b^\infty dB \frac{d\sigma(S)}{dB}, \]
decreases monotonically as $b$ increases from zero to infinity (see
Figure \ref{fg.centrality}). As a result, an experimentally determined
histogram of $f$ would determine $b$ uniquely, provided one knows the
functional form of $f(\sqrt S,B)$. This is not yet computable from QCD,
so one has to make models.

The simplest, and oldest, model is called the {\sl Glauber model\/}. In
this, one assumes that the nucleus-nucleus collision is described by
independent nucleon-nucleon collisions. The nucleons are distributed in
each nucleus according to the density determined by low-energy electron
nucleus collisions. Models which incorporate more phenomenology have
also been developed; see \cite{general} for more information. It has been
realized in recent years that the lumpy distribution of nucleons in the
initial state (see Figure \ref{fg.initene}) cannot always be averaged
over, but must be taken into account in these models.

\bef[h]
\begin{center}
\includegraphics[scale=0.1]{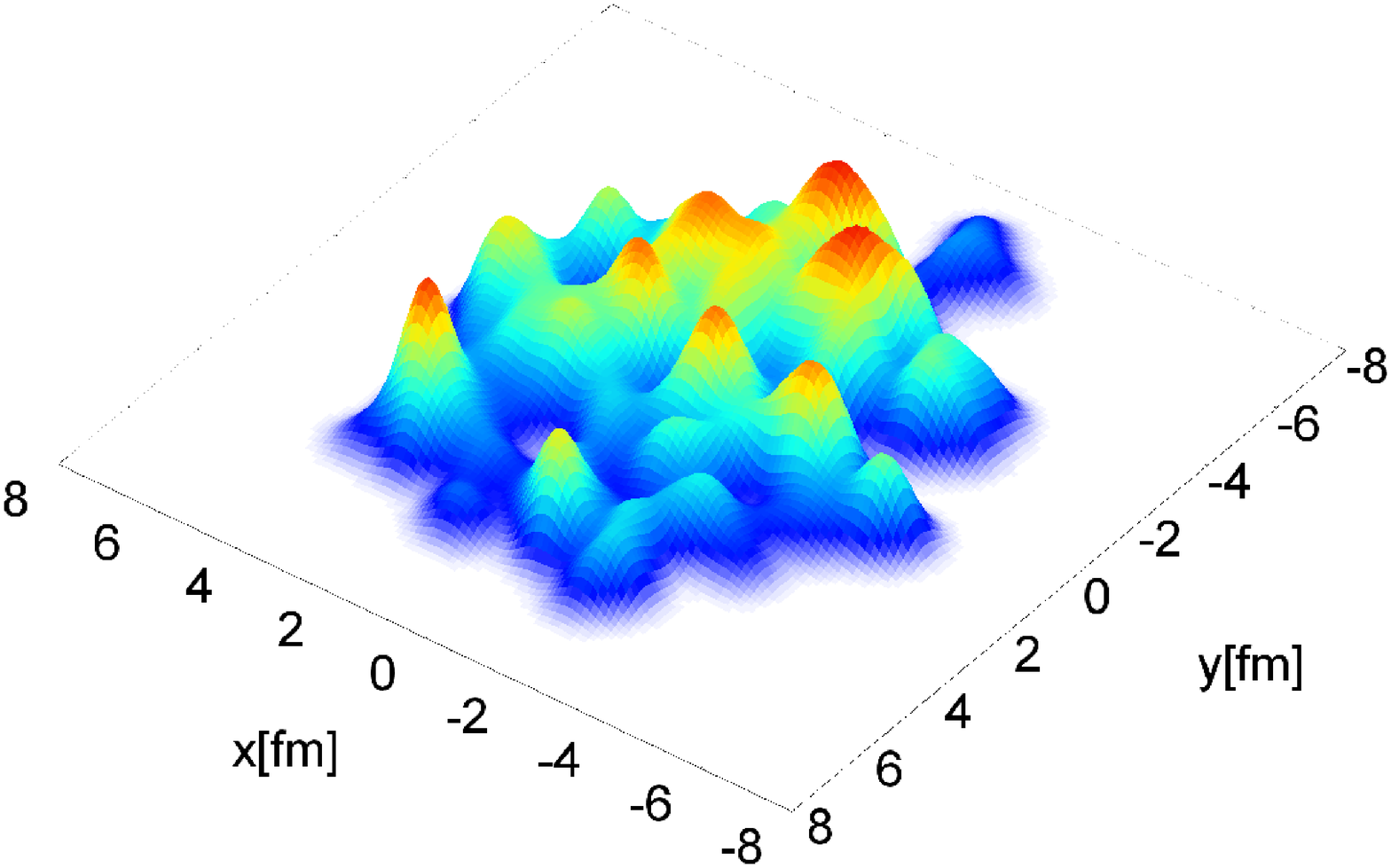}
\includegraphics[scale=0.1]{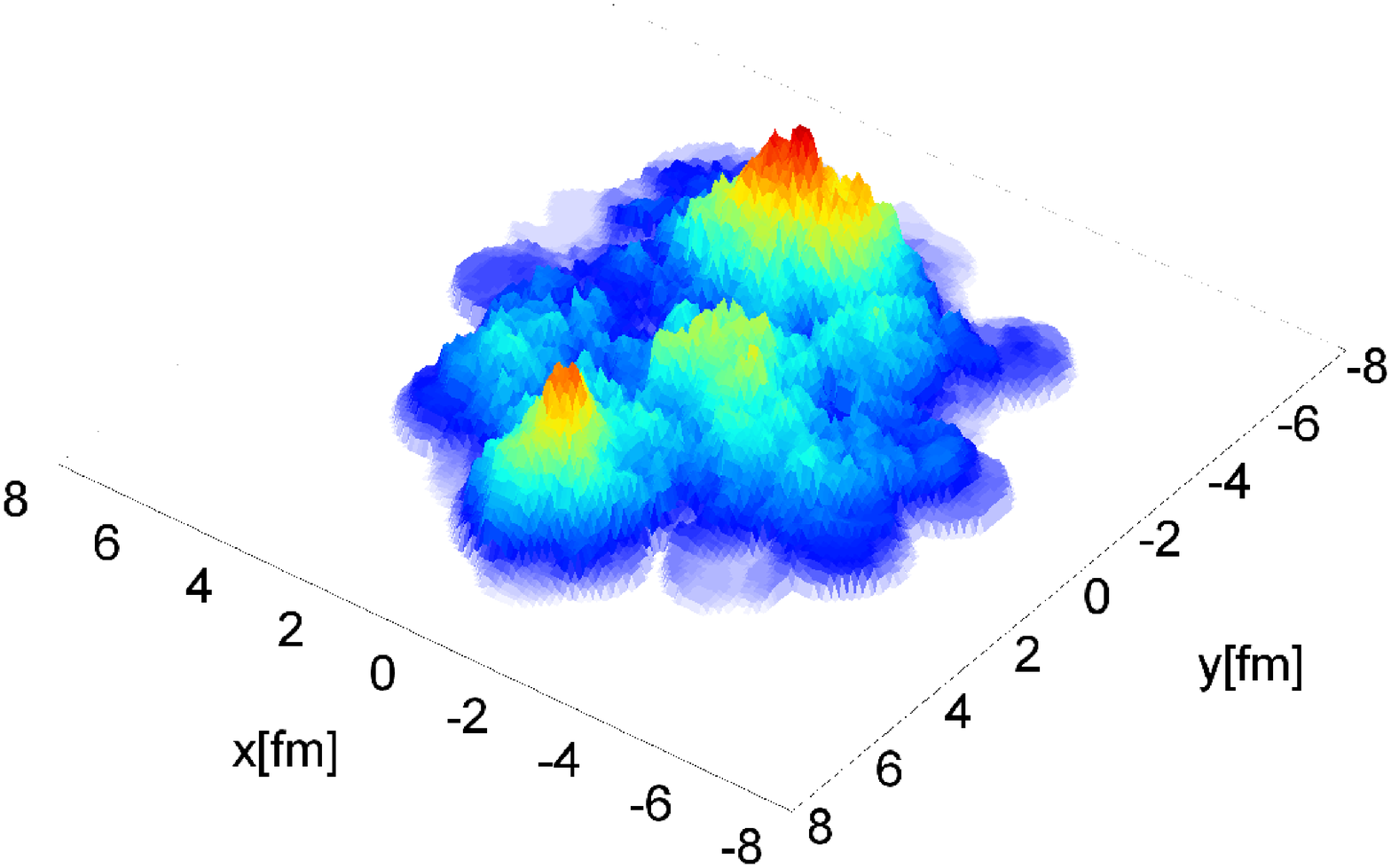}
\includegraphics[scale=0.1]{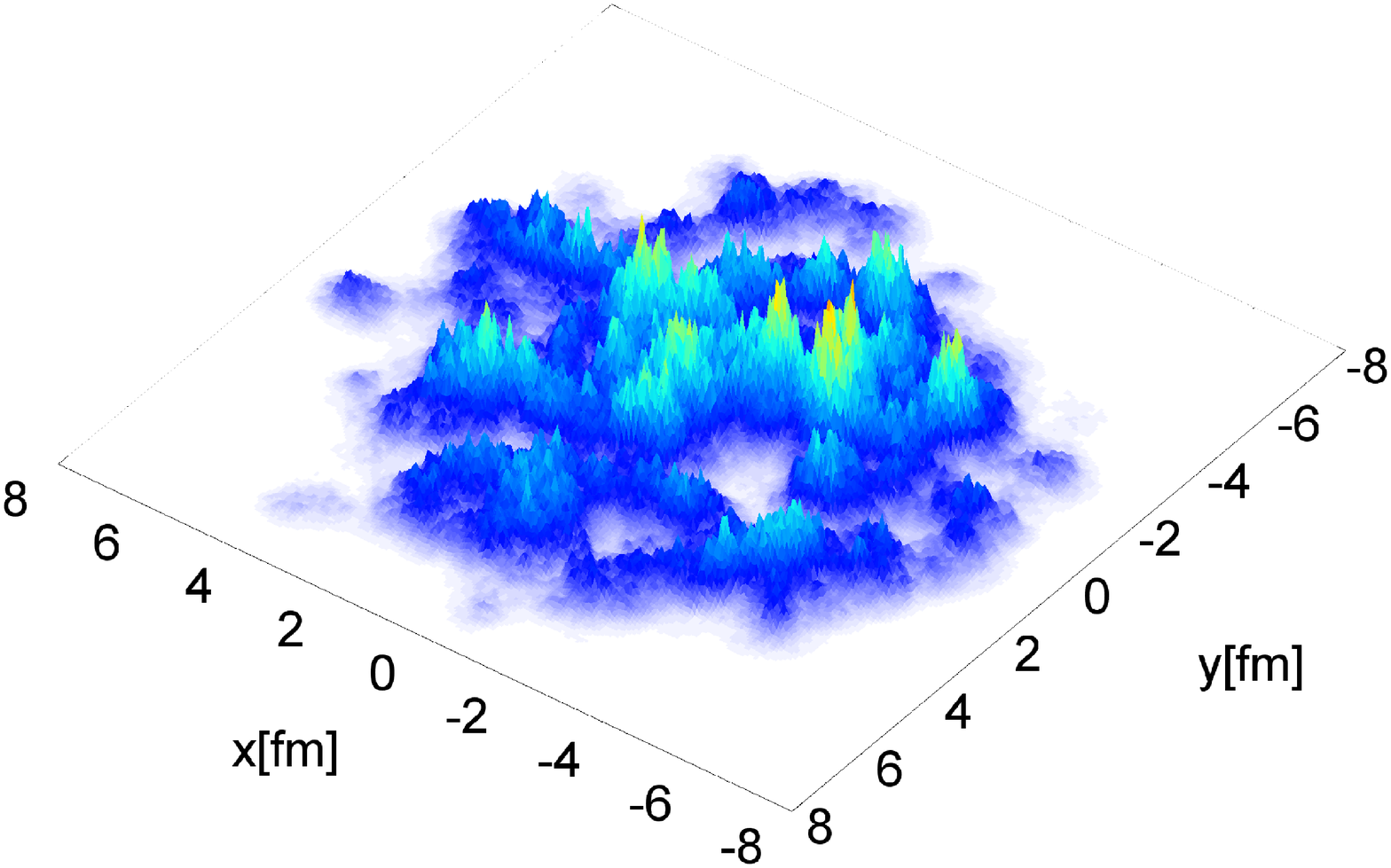}
\end{center}
\caption{Transverse energy profiles, 0.2 fm after the collision, in three
 models of initial states \cite{Gale1209}. The coarse-grained average of
 these distributions should give the nuclear density known through low-energy
 experiments. These relativistic experiments capture the quantum fluctuations
 in the initial nuclear wave-function.}
\label{fg.initene}\eef

\bef[ht]
\begin{center}
\includegraphics[scale=0.33]{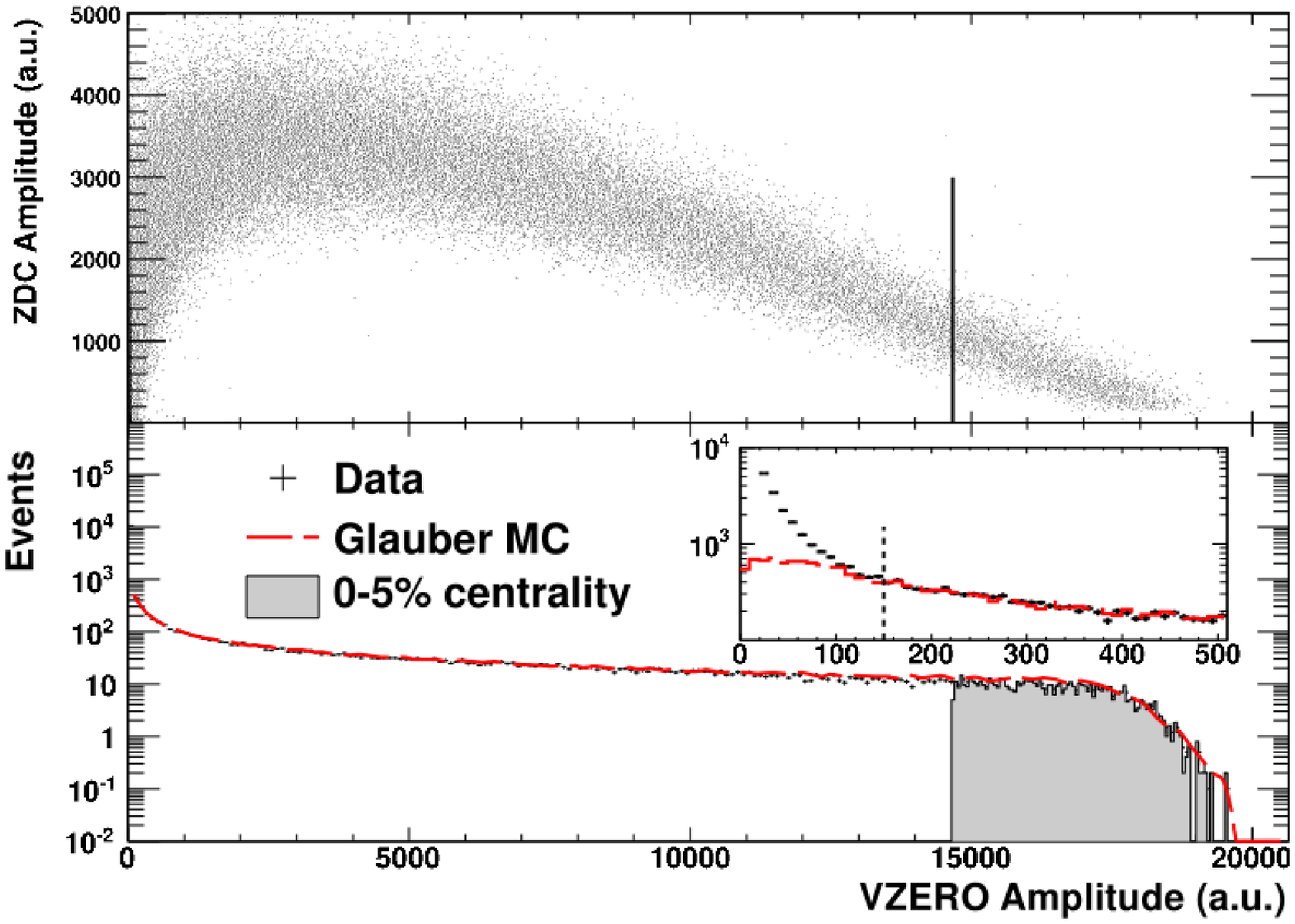}
\end{center}
\caption{One of the definitions of centrality used by the ALICE collaboration
 \cite{ALICE1011} uses the histogram of measurements in the VZERO module. The
 experiment determines its correlation with the energy deposition in the ZDC.}
\label{fg.alicecentral}\eef

As one sees, the connection between the impact parameter and the
percent of total cross section is indirect and model dependent. Also,
when one realizes that the positions of nucleons inside the nuclei
fluctuate from one event to another, it is clear that the notion of an
impact parameter, and even the size and shape of a nucleus, are merely
averaged quantities. For experimental purposes what is necessary is to
classify events according to the degree of centrality. For this it is
enough to define centrality by any measure which changes monotonically
with $b$; for example, by multiplicity, zero degree calorimetry,
\etc. Care is needed to relate these measures to each other through
careful analysis of the data. One such analysis is shown in Figure
\ref{fg.alicecentral}.

\section{Hard probes}

One thinks of the LHC as an arena of hard QCD, \ie, of processes which
convert the partons contained in protons into jets, heavy quarks, W/Z
bosons, hard $\gamma$, H and so on. The typical momentum scale in these
processes is of the order $Q\simeq\langle x\rangle\sqrt S\simeq500$ GeV.
Final state interactions are suppressed in pp collisions because of two
reasons. Firstly, the dense hadronic debris are separated from probes by
large angles, $\Delta\eta$. Secondly, the energy scale of any remaining
hadronic activity in the central rapidity region is small: $\langle
E_T\rangle \simeq \lqcd\simeq0.3$ GeV.

In heavy-ion collisions, the first argument can still be
supported. However, the second argument may fail if the number density
of particles, $n$, is large enough. Let us make an estimate by assuming,
as before, that $n=5/{\rm fm\/}^3$. We know that the actual value of $n$
at the LHC is larger, so our argument will be overly conservative. Assume
that the jet cone has radius\footnote{The radius of a jet cone is defined
to be $R=\sqrt{\Delta\eta^2+\Delta\phi^2}$ where we take $\Delta\eta$
and $\Delta\phi$ to be the jet opening angles.} $R=0.2$, and that it
travels about $\ell=10$ fm through the fireball of soft particles. Then
the net energy in the soft hadrons it can interact with is 
\beq
 {\cal E} \simeq \langle E_T\rangle n R \ell^3 \simeq 300 {\rm\ GeV\/},
\label{pedestal}\eeq
where we have made a conservative estimate that the average transverse
energy of the particles is $\langle E_T\rangle\simeq0.3$ GeV.  Since this
is comparable with the initial energy, final state interactions become
important. An interesting consequence which we discuss here is jet
quenching.

\bef[h]
\begin{center}
\includegraphics[scale=0.45]{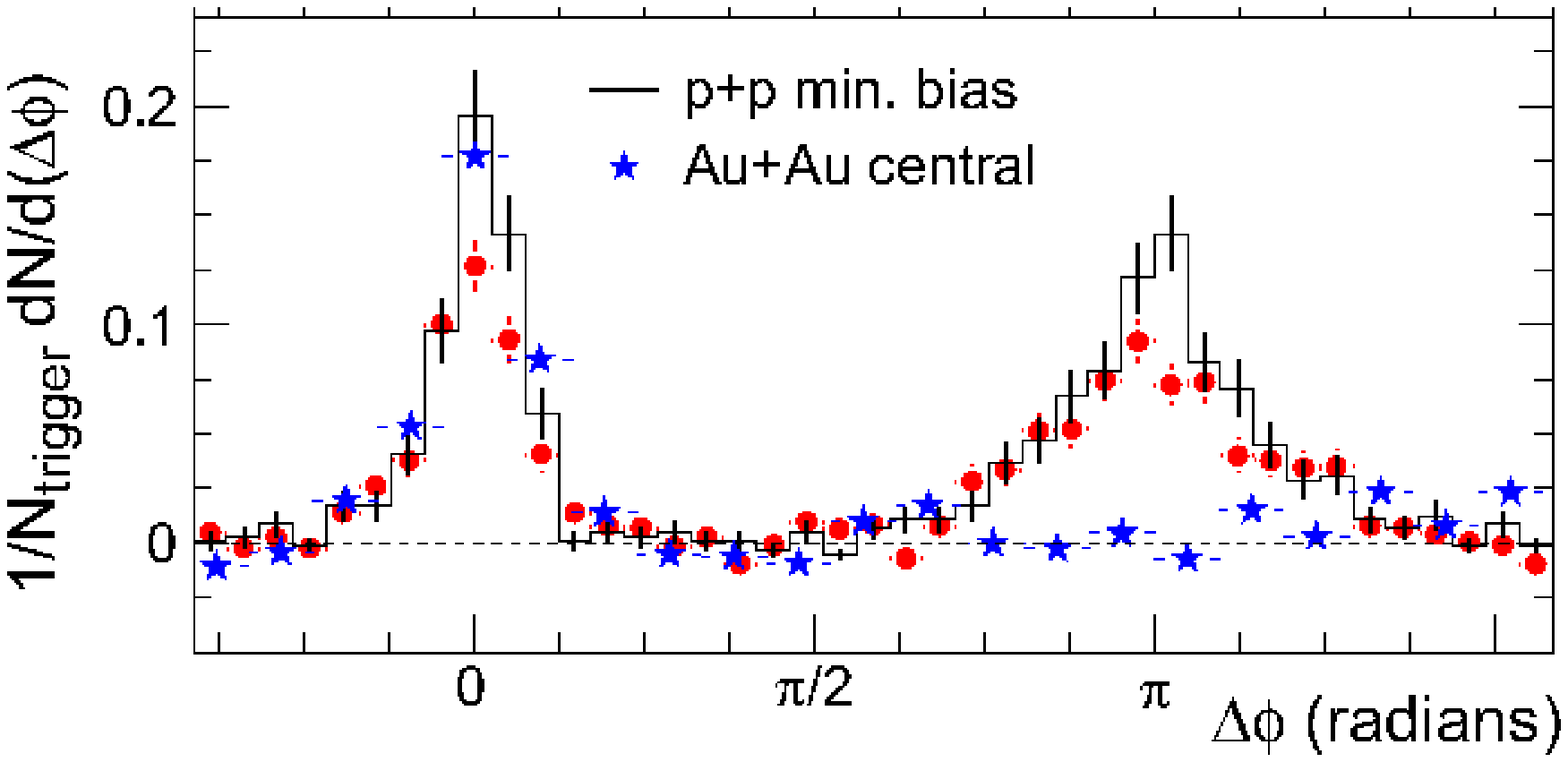}
\end{center}
\caption{Comparison of two-particle azimuthal distributions for central
 d+Au collisions (circles) to those seen in p+p (histogram) and Au+Au
 collisions (stars). The respective pedestals have been subtracted.}
\label{fg.jetq}\eef

When a jet evolves through a medium, interactions and radiation would
tend to deplete its energy \cite{jetq}. This simple idea is called
{\sl jet quenching\/}. The basic fact of jet quenching was beautifully
demonstrated by the STAR collaboration in BNL \cite{STAR0306} in the
plot given above. At $\sqrt S=200$ GeV jets are not very well developed,
and one must use high-$p_\T$ hadrons as proxies. STAR triggered on
events where there is a high-$p_\T$ hadron, and looked at the angular
distribution of the next highest-$p_\T$ hadron. In p+p collisions they
found a peak 180 degrees away (see Figure \ref{fg.jetq}). If the trigger
hadron can be assumed to come from a jet, then the backward peak comes
from an away-side jet which balances the momentum.  This was also seen in
d+Au collisions, thus demonstrating that initial state parton effects in
heavy nuclei do not wash away this peak. In Au+Au collisions they found
no peak in the backward direction: implying that the away-side jet is
hugely quenched\footnote{Since the near-side jet is used as a trigger,
the event sample is of those in which this is not completely quenched.}.

A measure of the quenching is provided by a comparison of the number of
jets of a given momentum in heavy-ion and proton collisions
\beq
   R_{AA}(b,y,p_\T) = \frac1{T_{AA}(b)} 
   \frac{d^3N_{AA}}{dbdydp_\T} \left( \frac{d^2N_{pp}}{dydp_\T} \right)^{-1}\,.
\label{raa}\eeq
Here $T_{AA}$ is an estimate of the number of proton pairs interacting
in AA collisions, and is usually extracted from a model, \eg, the
Glauber model. The numerator depends on collision centrality whereas the
denominator does not. Energy is tremendously more likely to flow from
the jet into the low-momentum particles in the medium (computations
reveal this in phase space factors). As a result, one would generally
expect $R_{AA}$ to be less than unity.

\bef[ht]
\begin{center}
\includegraphics[scale=0.45]{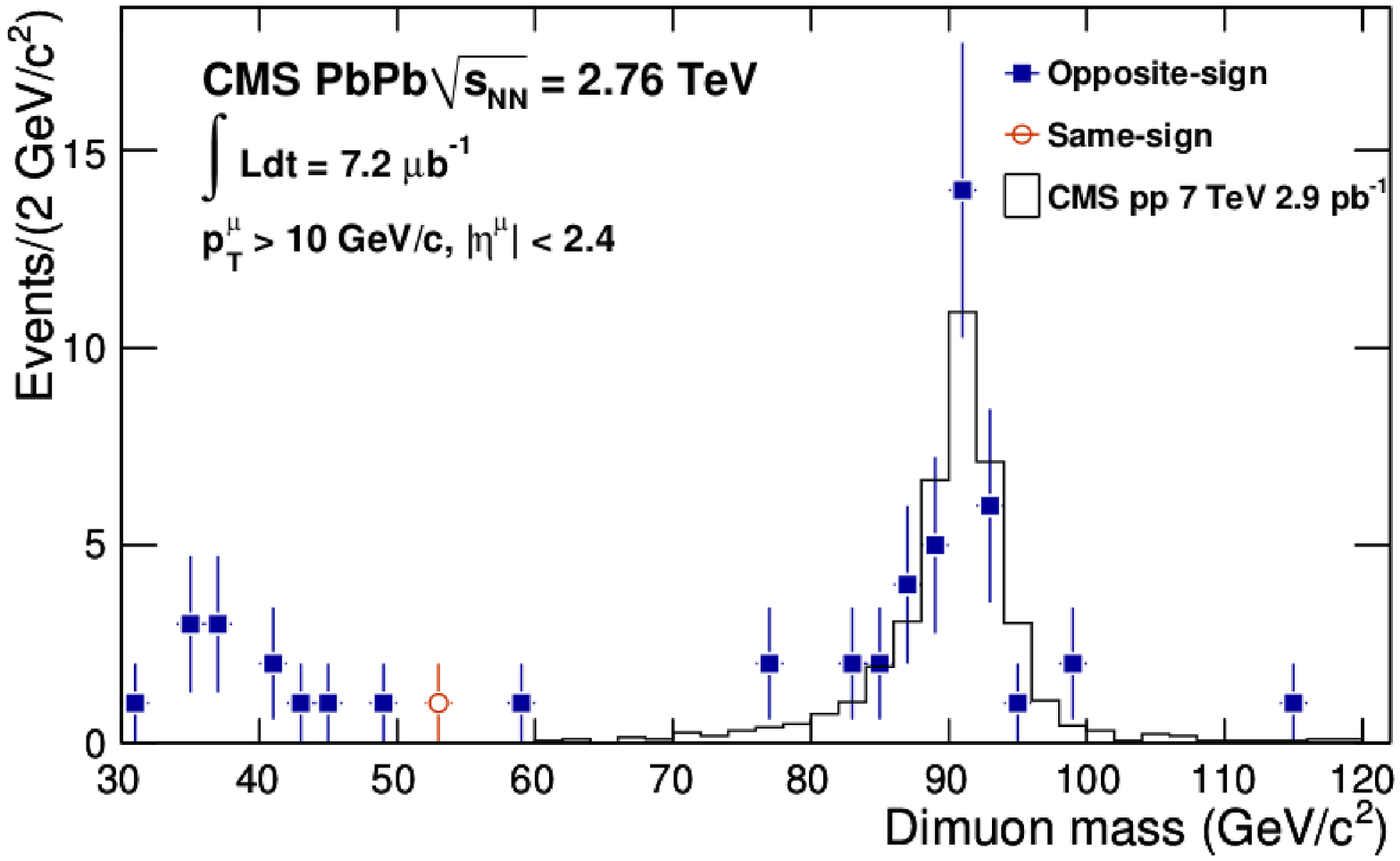}
\end{center}
\caption{The first attempt to constrain $T_{AA}$ from experiment.}
\label{fg.taa}\eef

Since a basic input into jet-quenching is $T_{AA}$, it is important to
constrain this through experiment. The production of high-$p_\T$ photons
or W/Z bosons provides this calibration. Since the vector bosons have
no strong interactions, the comparison of semi-inclusive single boson
production cross sections in pp and AA cross sections can directly measure
$T_{AA}$. One of the first attempts \cite{CMS1102} to constrain this
is shown in Figure \ref{fg.taa}. Small isospin corrections, shadowing,
and initial re-scattering effects must also be taken into account more
accurately in order to improve these constraints.

\bef[ht]
\begin{center}
\includegraphics[scale=0.45]{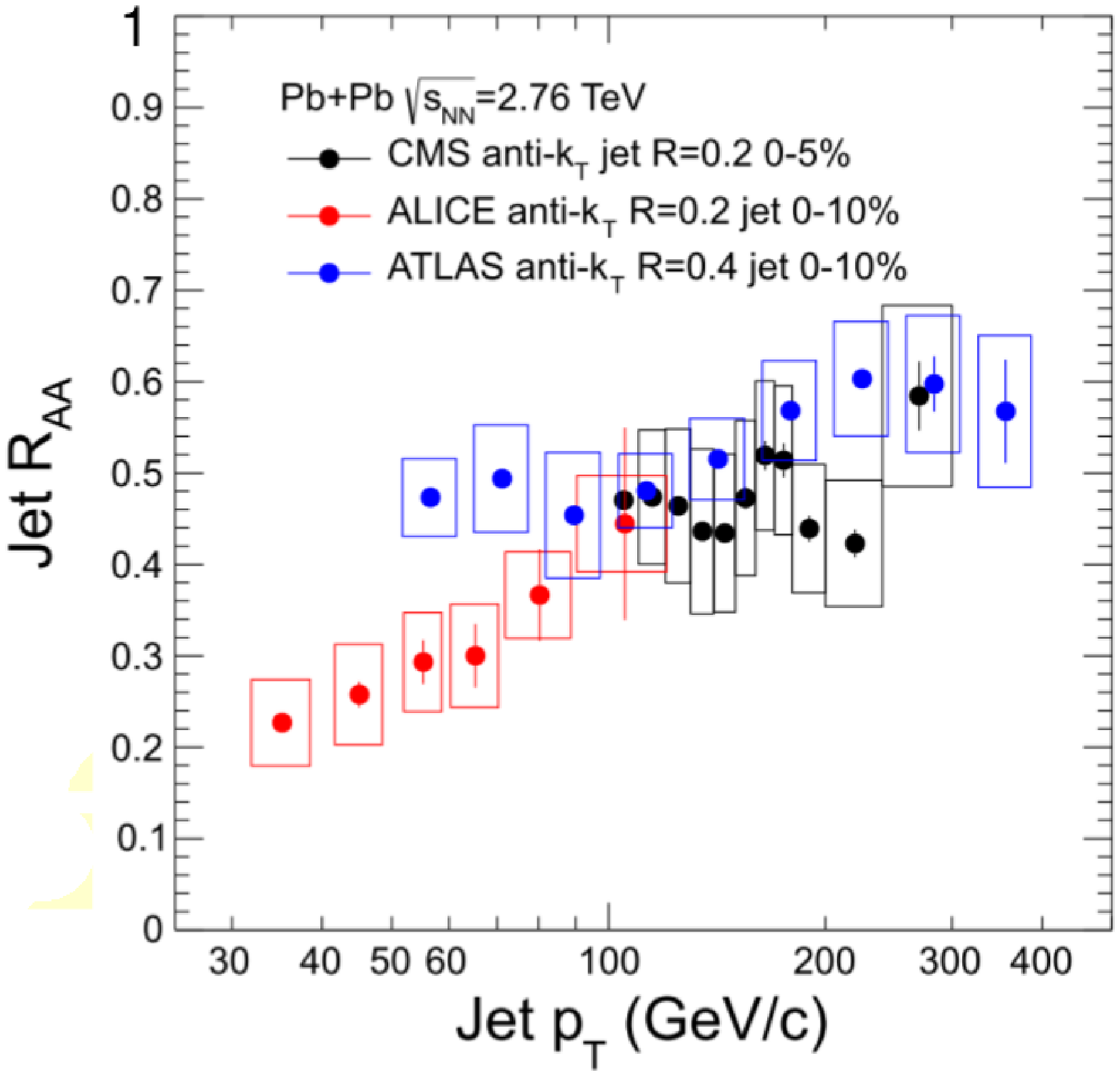}
\end{center}
\caption{A recent compilation of data on $R_{AA}$ \cite{Lee2014}.}
\label{fg.raadata}\eef


From the observations of $R_{AA}$ (see Figure \ref{fg.raadata})
one can extract a measure of the $p_T$ change of the jet per unit
distance travelled within the plasma: \[ \hat q = \frac1L\int
\frac{d^2p_T}{(2\pi)^2}\,p_T^2\,P(p_T,L). \] Most attempts to extract
this from data give $\hat q\simeq$ 1--2 GeV${}^2$/fm, \ie, in the range
of interest, $\hat q/T^3\simeq$4--5.  One should be able to extract
this number for QCD, but it turns out to be a vexing problem. There
are two main methods to handle problems in QCD: perturbative QCD is
used to compute processes where all momenta involved are large, and
lattice QCD provides a tractable computational method when all momenta
are small. Jet quenching couples a large momentum object (the jet) to
low-momentum objects (the medium). Nevertheless there have been attempts
to compute this in QCD using weak-coupling expansions \cite{Baier} or,
more recently, lattice QCD \cite{Majumder}. There are also computations
in cousins of QCD which have ${\cal N}=4$ supersymmetry in the limit of
large $N_c$ \cite{Wiedemann}.

$R_{AA}$ is just the simplest of experimental variables which can be
constructed. In order to understand how the medium steals energy and
momentum from the jet one should also understand medium modification of
rapidity and angular correlation, momentum imbalance between reconstructed
jets,  fragmentation functions, and jet substructure.

%
%
%
%
%

\section{Flow}

Once the fireball reaches local thermal equilibrium a much slower process
begins of transport of energy, momentum, and other conserved quantities
through the fireball. This is the hydrodynamic regime \cite{bj}. Tests
of hydrodynamics involve the study of quantities which are called flow
coefficients \cite{Ollitrault}. In order to understand what these are, we
need to think again about the geometry and kinematics of the collisions.

In the collision of point-like particles, there is a rotational
symmetry around the beam axis. As a result, cross sections or particle
production rates depend only on the scattering angle $\theta$ (or
equivalently, on $\eta$ or the rapidity $y$) and the transverse momentum,
$p_\T$. Kinematically, there is only one initial vector in the center of
mass (CM) frame of the problem, the initial momentum $\V k$ of one of the
particles (the other particle has momentum $-\V k$). Final state momenta
see only the angle from $\V k$, which is $\theta$, and the transverse
projection $p_\T$.

In heavy-ion collisions, there is a second initial vector: $\V b$, which
is the line between the centers of the nuclei. The existence of such
a vector, not collinear with $\V k$, means that the azimuthal symmetry
around $\V k$ is broken in the initial state, and final state momentum
distributions may depend on angles the final momentum makes with both
$\V k$ and $\V b$ as well as $p_\T$. Conventionally, these distributions
are given in terms of $\eta$, $p_\T$, and the azimuthal angle $\phi$.
The two vectors $\V k$ and $\V b$ lie on a plane which is called the
{\sl reaction plane}. This breaking of cylindrical symmetry also occurs
in proton-nucleus collisions, since the proton can meet the nucleus with
a non-vanishing impact parameter. In very high energy collisions, the
increasing proton-proton cross section implies that the swollen protons
can also be treated similarly. One may already be seeing such effects
in the sample of extreme high multiplicity events in pp collisions at
the LHC.

The {\sl flow coefficients} are the Fourier transforms of velocity
distributions with respect to $\phi$ \cite{Voloshin}. The $n$-th
Fourier coefficient is denoted by the symbol $v_n$. These are normally
taken at $y=0$ not only because of the limited rapidity coverage of
heavy-ion detectors, but also because one expects the fireball to be
well-separated from the fragmentation region. Nevertheless, studying
the rapidity dependence of flow coefficients is of some interest. The
study of the $k_\T$ dependence of the $v_n$ is of great interest.

\bef[th]
\begin{center}
\includegraphics[scale=0.65]{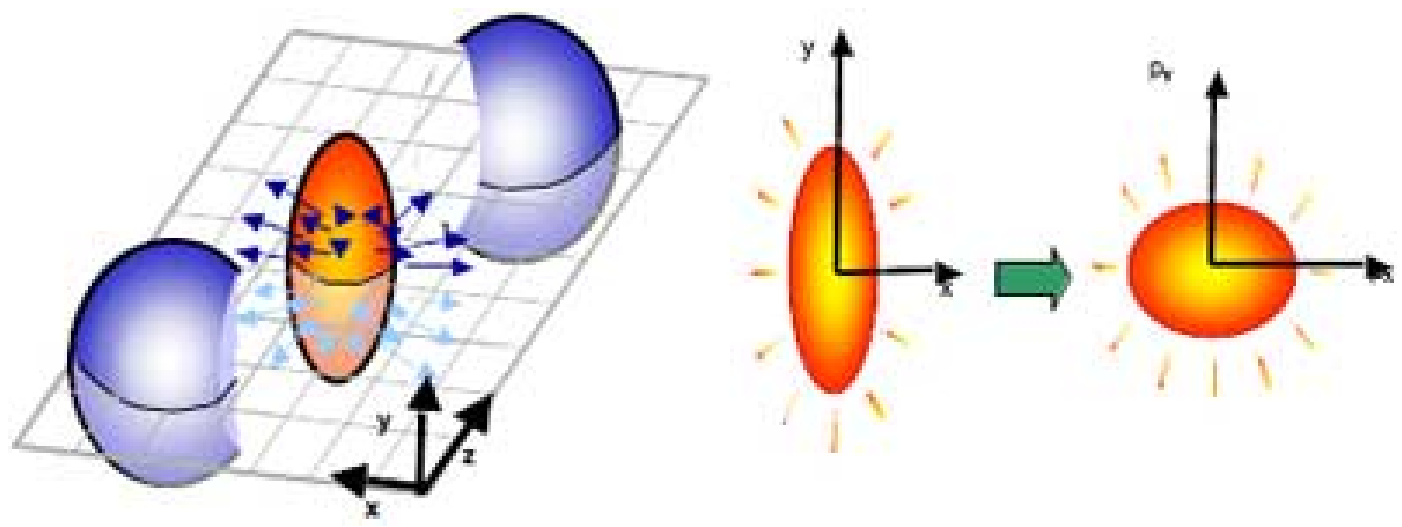}
\end{center}
\caption{The geometry of elliptic flow.}
\label{fg.aniso}\eef

Clearly, the reaction plane in different collisions can rotate around
the beam axis, so single particle distributions will recover azimuthal
symmetry when averaged over events.  Although the overall orientation
of the reaction plane is forgotten on the average, the relative angles
between two particles remembers the difference from the reaction plane.
So, in order to see the flow coefficients one has to construct the
angular correlations of two or more particles.

In the collision of symmetric nuclei, $\V b$ and $-\V b$ seem to be
completely equivalent. As a result the two sides of the reaction plane
seem to be completely symmetric. This implies that only the even flow
coefficients, $v_2$, $v_4$, \etc, are non-vanishing ({\sl elliptic
flow\/} is the name given to $v_2$). However, when one studies the flow
event-by-event (E/E) one has to take into account the fact that the
positions of nucleons inside the nuclei may fluctuate. Then there may be
more nucleons on one side of the plane, so breaking this orientational
symmetry around the reaction plane, as a result of which odd harmonics
may exist. Currently there are studies of the {\sl directed flow\/}
$v_1$, {\sl triangular flow\/} $v_3$, and even the coefficient $v_5$.
The flow coefficients yield a combination of information on the initial
state and the evolution of fireball. E/E fluctuations of flow coefficients
yield more refined information on the initial state \cite{Srivastava}

It is claimed that the observations of $v_2$ imply the formation of
locally thermalized matter in heavy-ion collisions. Although this argument
is technical it is easy to understand this intuitively. In the off-center
collisions of nuclei the colliding region is a pellet. Particles formed
in the initial collisions have distributions which have positional
anisotropy, $\epsilon_n$, the pellet being long in one direction (see
Figure \ref{fg.aniso}). The generation of $v_n$ involves transforming
$\epsilon_n$ into momentum anisotropy.  This is impossible unless there
is hadronic re-scattering.  Also, the measurements of $v_2$ show that
the momentum is larger in the direction in which the original position
distribution was squeezed. This is hydrodynamic flow, since that is
driven by pressure gradients, and the gradients in the shorter direction
are larger.

\bef[h]
\begin{center}
\includegraphics[scale=0.50]{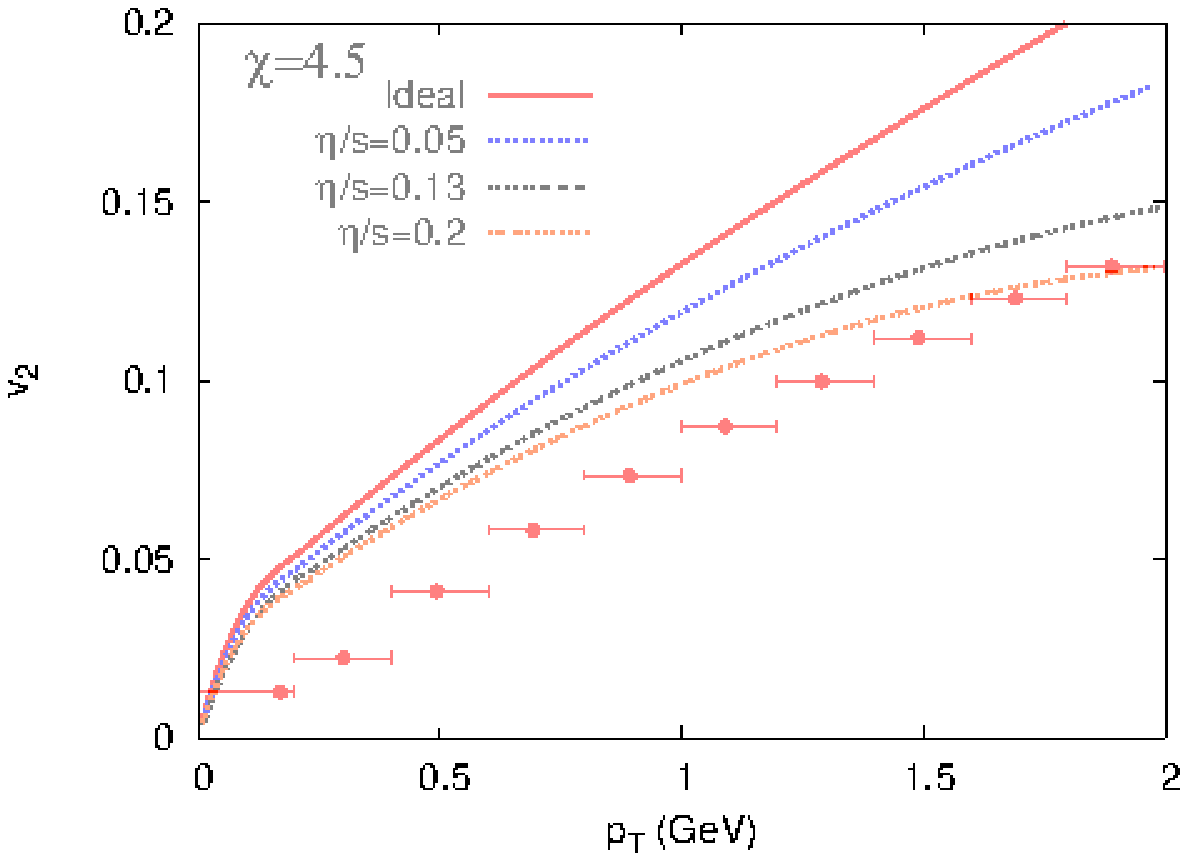}
\end{center}
\caption{Predictions for elliptic flow from ideal hydrodynamics compared
 to data. See the text for a discussion.}
\label{fg.teaney}\eef

The technical question here is how well does hydrodynamics explain
the observed $v_2$. Ideal hydrodynamics, \ie, hydrodynamics without
dissipation, is a toy model which is often used to understand general
features of data.  This already come within a factor two of the
data, and shows the same $p_\T$ dependence as the observations (see
Figure \ref{fg.teaney}). It also fails in the ``right'' direction,
in that it over-estimates $v_2$. Dissipation would clearly reduce
the predictions, and bring it closer to observations \cite{Dumitru}.
One of the first results (Dusling and Teaney in \cite{Dumitru}) is shown
in the figure. Intense work continues to be done in understanding the
implications of the data and the constraints from QCD \cite{flow}.



\section{Chemical composition of the final state}

The most easily observed quantities have to do with the final
state. The basic observables are the spectra of identified particles. The
multiplicity of each type of particle, $\pi^\pm$, $K^\pm$, \etc, is called
its yield. This is the integral over the spectrum.  Relative yields
of hadrons is the outcome of {\sl hadron chemistry\/}, \ie, inelastic
re-scattering in the final state. Examples are,
\beq
   p + \pi^- \leftrightarrow n + \pi^0, \qquad
   p + \pi^- \leftrightarrow \Lambda + K^0,
\label{reacs}\eeq
The rates of such predictions determine whether hadron chemistry comes to
chemical equilibrium.

As the fireball evolves, eventually mean-free paths or relaxation times
become comparable to the size or expansion rate. When this happens, local
thermal equilibrium can no longer be maintained, and hydrodynamics cannot
be supported. Then the components of the fireball are said to {\sl freeze
out\/}.  In principle freeze out could occur either before the fireball
cools into hadrons or after. Under normal circumstances, \ie, if the
thermal history of the fireball does not take it near a phase transition,
then it seems to freeze out in the hadronic phase. This implies that the
later stages of hydrodynamics require the equation of state of a hadronic
fluid. Since hadrons are massive, inelastic collisions, \ie those which
change the particle content of the fireball, require larger energies than
elastic collisions. As a result, {\sl chemical freezeout\/}, \ie, fixing
of the hadron content of the fireball, may occur earlier than {\sl kinetic
freezeout\/}, \ie, fixing of the phase space distribution of hadrons.

In the fireball the particles interact strongly enough that a temperature
is maintained.  However, at freeze out the interactions stop abruptly. So
all hadrons emitted by the fireball at freeze out can be assumed to be an
ideal gas of particles coming from a source whose temperature is set at
the freeze out. This simple approximation, which goes by the name of the
{\sl hadron resonance gas model\/} has had remarkable phenomenological
success \cite{Cleymans}. However, recent measurements at the LHC (ALICE
collaboration \cite{Cleymans}) and a more careful look at RHIC results
shows interesting discrepancies which imply that this model needs to
be improved.

At early times, the fireball is a reactive fluid whose description
requires coupling of hydrodynamics with diffusion and flavour
chemistry. The reaction rates depend on local densities as well as
rates of mixing due to fluid movement, known as advection, as well as
diffusion. In order to make quantitative predictions, one must first
understand whether advection or diffusion is more important in bringing
reactants together. This is controlled by {\sl Peclet's number\/}
\beq
 \Pe = \frac{Lv}D = \frac{Lv}{\xi c_s} = {\rm Kn\/}\,M,
\label{peclet}\eeq
where $L$ is a typical macroscopic distance within the fireball over
which we wish to compare advection and diffusion, $v$ a typical flow
velocity, $\xi$ is a typical density-density correlation length and
$c_s$ is the speed of sound. The diffusion constant, $D\simeq\xi c_s$.
We have also used the notation for the Mach number of the flow, $M=v/c_s$,
and the Knudsen number, ${\rm Kn\/}=L/\xi$.  When $\Pe\ll1$ diffusion
is more rapid than advection; when $\Pe\gg1$ advection is more rapid
\cite{Bhalerao0901}.

Peclet's number defines a new length scale in the fireball, this is the
scale at which advection and diffusion become comparable---
\beq
 L \simeq \frac\xi M.
\label{peclength}\eeq
Since longitudinal flow has $M\le\sqrt 3$, then taking $\xi$ to be
approximately the Compton wavelength of a particle, we find that for
baryons, $L\simeq0.3$ fm and for strange particles, $L\simeq0.5$ fm.
This implies that advection may be important in chemical processes
occurring in the early stages of the evolution of the fireball, but
over most of its history, the availability of reactants is governed
by diffusion.

Once the reactants have been brought together we can ask whether one or
the other reaction channel is available. If the reactions are slower
than the time scale of transport, then we may consider the fireball
to be constantly stirred. It is then enough to examine chemical rate
equations. In this approximation, a toy model which takes into account
only pion and nucleon reactions is:
\begin{eqnarray}
\nonumber
   \dot p &=& -\gamma(p\pi^0 -n\pi^+) -\gamma'(p\pi^--n\pi^0)+\cdots,\\
\nonumber
   \dot n &=&  \gamma(p\pi^0 -n\pi^+) +\gamma'(p\pi^--n\pi^0)+\cdots,\\
\nonumber
   \dot\pi^0 &=& -\gamma(p\pi^0 -n\pi^+) +\gamma'(p\pi^--n\pi^0)+\cdots,\\
\nonumber
   \dot\pi^+ &=&  \gamma(p\pi^0 -n\pi^+)+\cdots,\\
\nonumber
   \dot\pi^- &=& -\gamma'(p\pi^--n\pi^0)+\cdots.
\label{isomodel}\end{eqnarray}
Here the label for a particle denotes the density of that particle.
The rate constants $\gamma$ and $\gamma'$ can be deduced from experimental
measurements of cross sections. The equilibrium concentrations are given by
\beq
 \frac pn=\frac{\pi^+}{\pi^0}=\frac{\pi^0}{\pi^-} \quad(=\zeta),
\label{model}\eeq
where $\zeta$ is the isospin fugacity. Since $\pi^+/\pi^-=\zeta^2$,
if we set $\zeta\simeq1$, then $\mu_I=T\log\zeta\simeq0$. Even in this
simple limit of a very rapidly stirred fireball, a more realistic model
contains all possible reactions between many species of particles, of
which many cross sections are unmeasured. As a result, a detailed model
is out of reach and one must develop simplified models which catch as
much of the physics as the state of the data justifies.

\bef[h]
\begin{center}
\includegraphics[scale=0.45]{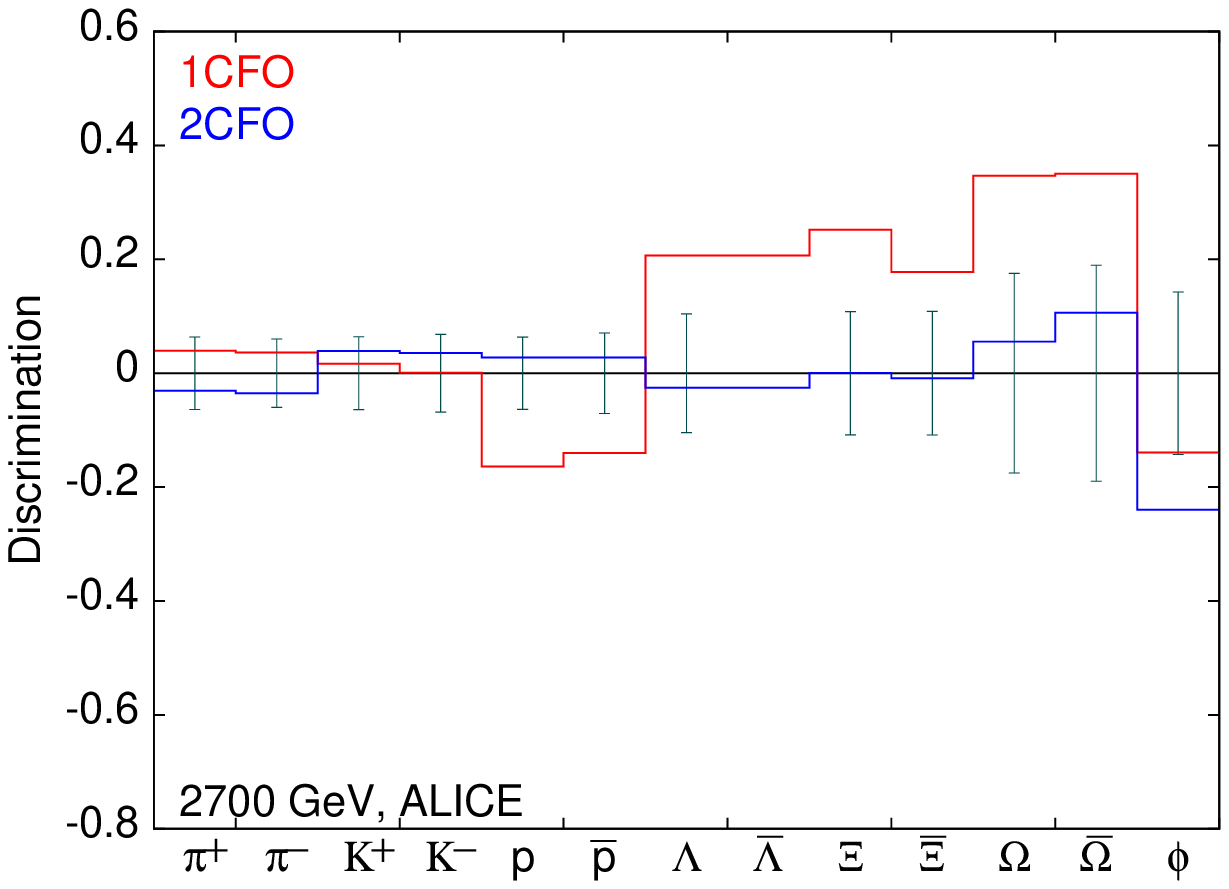}
\includegraphics[scale=0.45]{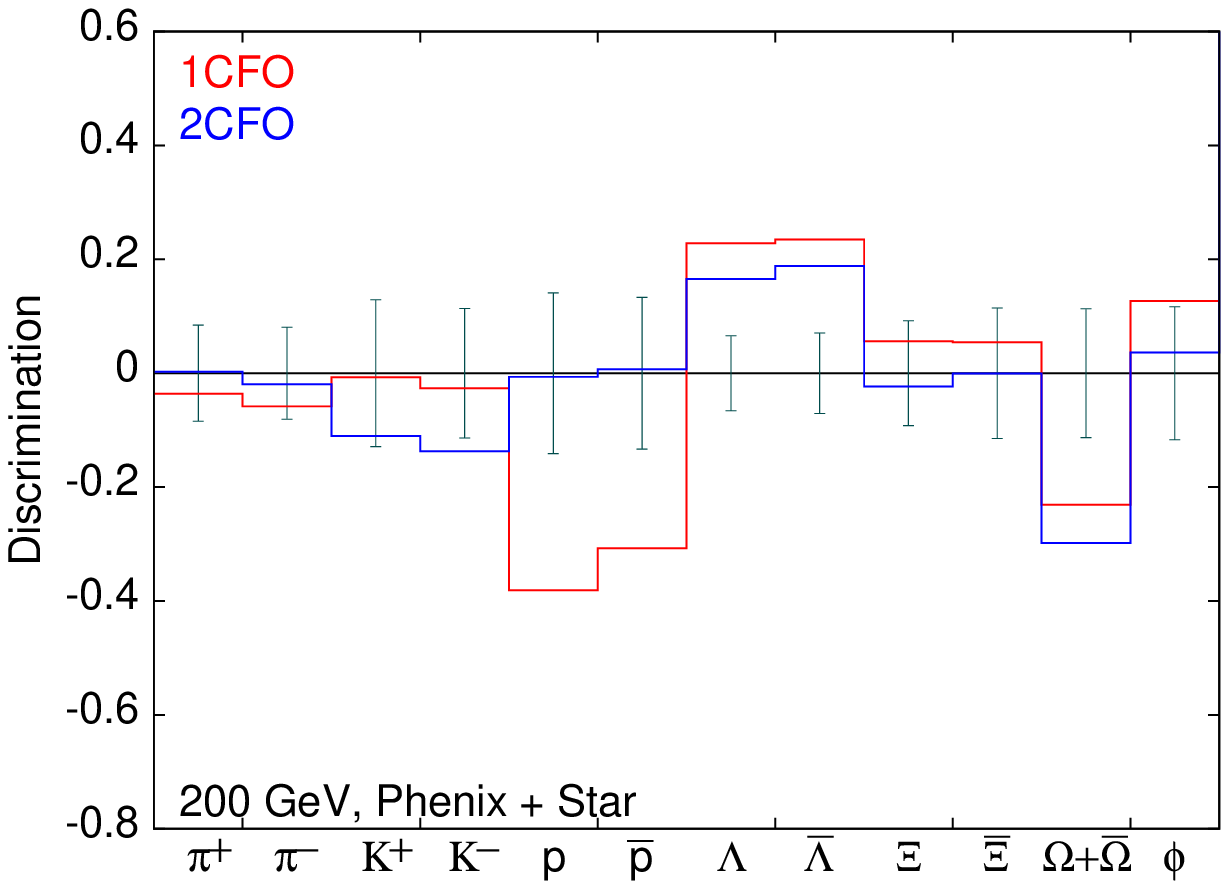}
\end{center}
\caption{Comparison of model predictions with data in terms of the discriminant
(model-data). The closer this is to zero, the better the model. The error bars
show the error on the data and set the scale of what is acceptable mismatch
between data and model. At all energies, 2CFO works better than 1CFO.}
\label{fg.tcfo}\eef

\bef[ht]
\begin{center}
\includegraphics[scale=0.65]{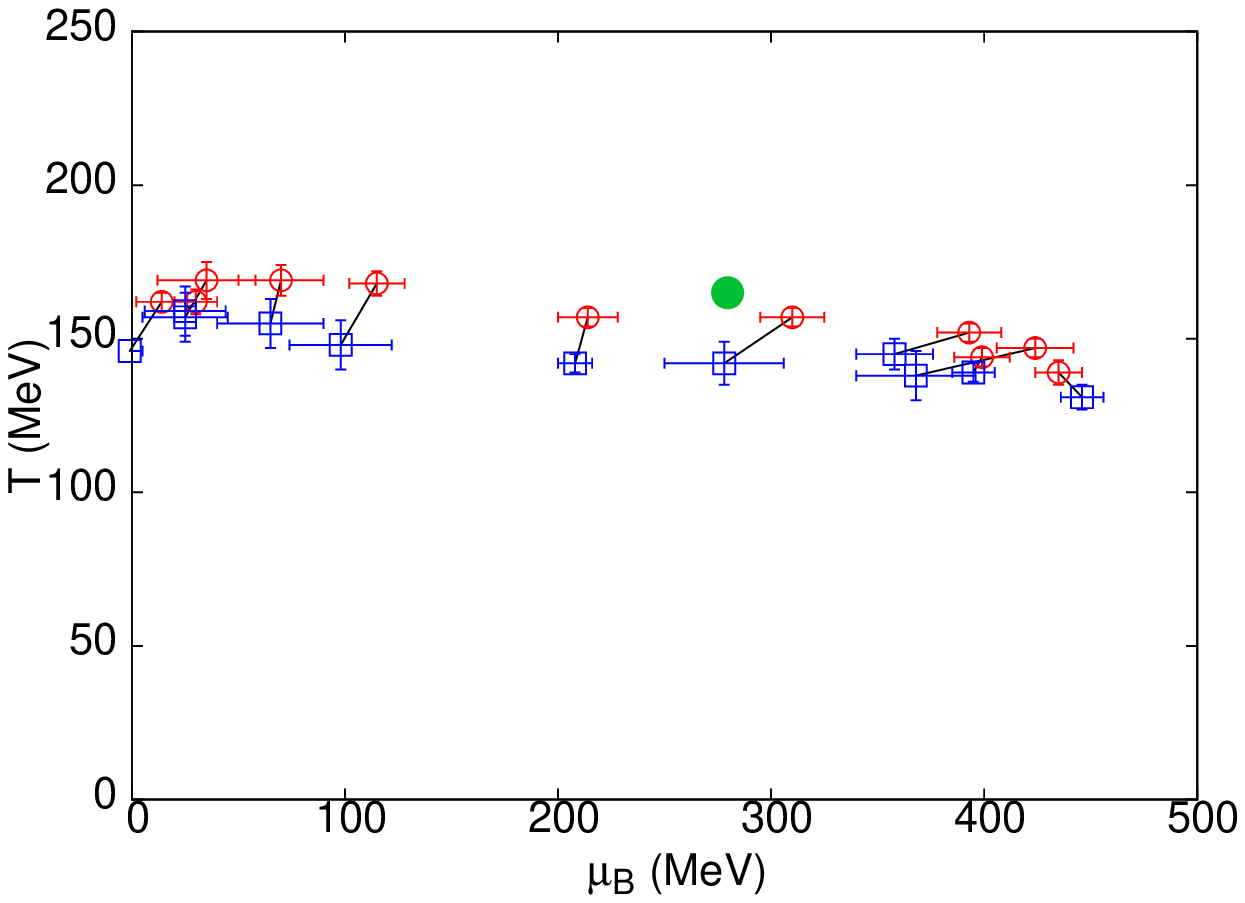}
\end{center}
\caption{The freeze out points obtained in 2CFO (from Chatterjee \etal,
 \cite{2cfo}). The strange freezeout point is shown with circles, and
 the non-strange with squares. The two freeze out points at the same
 $\sqrt S$ are joined with a line. The large filled circle is the 
 estimated location of the QCD critical point from lattice computations
 \cite{ilgti}.}
\label{fg.fos}\eef

In order to set up such a model, we consider flavour
changing reactions. Strangeness changing processes seem to naturally split
into subgroups. {\sl Indirect transmutations\/} of $K$ and $\pi$ involve
strange baryons in reactions such as $\Omega^- + K^+ \leftrightarrow
\Xi^0 + \pi^0$. These have very high activation thresholds.  {\sl Direct
transmutations\/} can proceed through the strong interactions such as
$K^+ + K^- \leftrightarrow \pi^+ + \pi^-$. These are OZI violating
reactions; slower than generic strong-interaction cross sections.
Direct transmutations through weak interactions are not of relevance in
the context of heavy-ion collisions. As a result, there is no physics
forcing $K$ and $\pi$ to freezeout together. However $K$ and $\phi$ are
resonantly coupled, so they may freeze out together \cite{2cfo}.  On the
other hand, isospin changing processes (the model in eq.\ \ref{isomodel})
require extremely low activation temperatures, and may persist till later.

One can capture this information into a HRG model with two freezeout
points: one for the strange hadrons and $\phi$ (since this is resonantly
coupled to the $K^\pm$ channel), another for non-strange hadrons. We
could call this model 2CFO \cite{2cfo} in contrast with the usual HRG
model with a single freeze out (1CFO). A comparison of measurements and
best fit model predictions is shown in Figure \ref{fg.tcfo}.

Interestingly, the introduction of two freeze out points allows one to do away
with some unphysical features of the freeze out model 1CFO. In most such models
there is a mismatch between strange and non-strange baryon production, which
is fixed by having a fugacity factor which changes the occupancy of strange
hadrons. This factor cannot be justified within an ideal gas picture, nor
does it vary smoothly or monotonically as $\sqrt S$ is changed. Such nuisance
parameters no longer appear within the 2CFO scheme.

The success of the 2CFO scheme implies that as one introduces more of the
hadron dynamics into the freezeout process, the ability to describe the data
improves. This justifies our belief that a proper description of reactive
transport should be able to give a good description of the final observed
yields.

The freeze out temperatures and chemical potentials in 2CFO are shown in
Figure \ref{fg.fos}. Also shown there is the position of the critical point
of QCD determined in lattice studies \cite{ilgti}. The freeze out curves
pass close to the QCD critical point, making it plausible that a study of
the final state as one scans in $\sqrt S$ can reveal signals of this very
interesting prediction of QCD. This is the rationale for the RHIC Beam
Energy Scan (BES) program and for planned future experiments in GSI and JINR.

Experiments also measure the yields of heavy-quarkonia. In particular, the 
yield of the $\Upsilon$ family of mesons in AA collisions at LHC differs
significantly from that in pp collisions at same $\sqrt S$. This is usually
reported in terms of an $R_{AA}$ for the meson. Since the quark mass is
large, $M\gg T\simeq\lqcd$, one may expect that the production of
quarkonia is a hard process. However the binding energy is of the order of
the temperature, $B\simeq T$, so we may expect large thermal effects as the
cause of the change between AA and pp collisions \cite{matsui}.

\bef[ht]
\begin{center}
\includegraphics[scale=0.65]{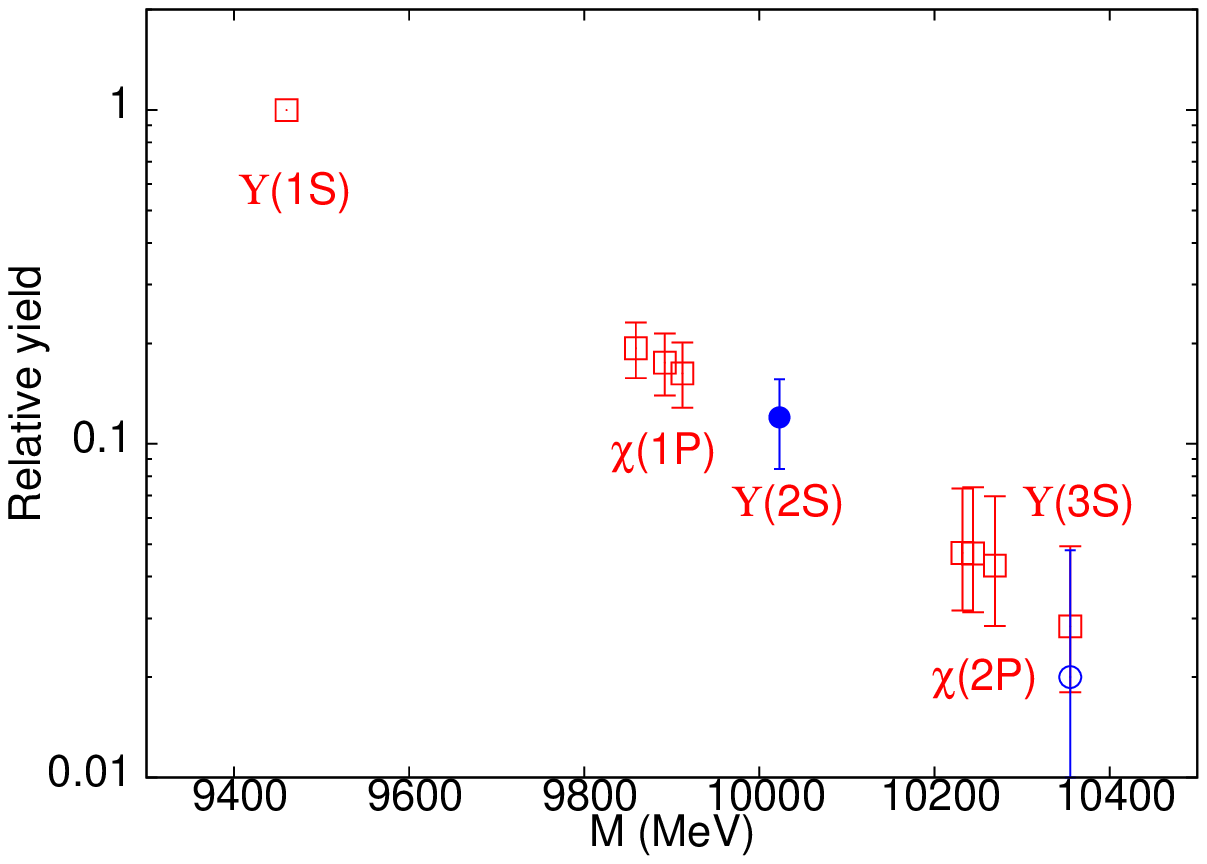}
\end{center}
\caption{Measured (filled circles) and predicted (unfilled squares)
 suppression in the bottomonium family using a simple thermal model
 for freeze out \cite{Sharma1401}.}
\label{fg.upsilon}\eef

Thermal lattice QCD computations show that the highest mass resonances,
which are the least bound, are more easily disrupted at any given
temperature \cite{jpsilat}. This observation led to the formulation of
a key observation called {\sl stepwise suppression\/}, \ie, as $\sqrt
S$ is increased $R_{AA}$ of the higher resonances drop below unity,
roughly in the order of the binding energy \cite{stepwise}. If this
works, then at a sufficiently high temperature it should be possible to
use a thermal model to understand the relative yields of the $\Upsilon$
family of mesons using the variables
\begin{equation}
 r[\Upsilon(n\ell)] = \frac{dN_{AA}^{\Upsilon(n\ell)}}{dydp_T}
    \left(\frac{dN_{AA}^{\Upsilon(1S)}}{dydp_T}\right)^{-1}.
\label{rjpsi}\end{equation}
The thermal model involves only a single parameter: the freeze out
temperature of this family of mesons. The first data 
from the LHC \cite{upsdat} is fitted \cite{Sharma1401} well by
\begin{equation}
   T_f^\Upsilon = 222^{+28}_{-29} {\rm\ MeV}.
\label{tfrupsilon}\end{equation}
It will be interesting in future to see whether other members of
the bottomonium family confirm this picture. Future data on
$r[\Psi(n\ell)]$ will also provide useful tests. More detailed
dynamical models \cite{jpsidyn} predict many more details of the
kinematics of quarkonium suppression.

\section{Fluctuations}

Is the ensemble of heavy-ion collision events captured by a detector
related to the ensembles required to study the thermodynamics of strong
interactions? 

\bef[t]
\begin{center}
\includegraphics[scale=0.4]{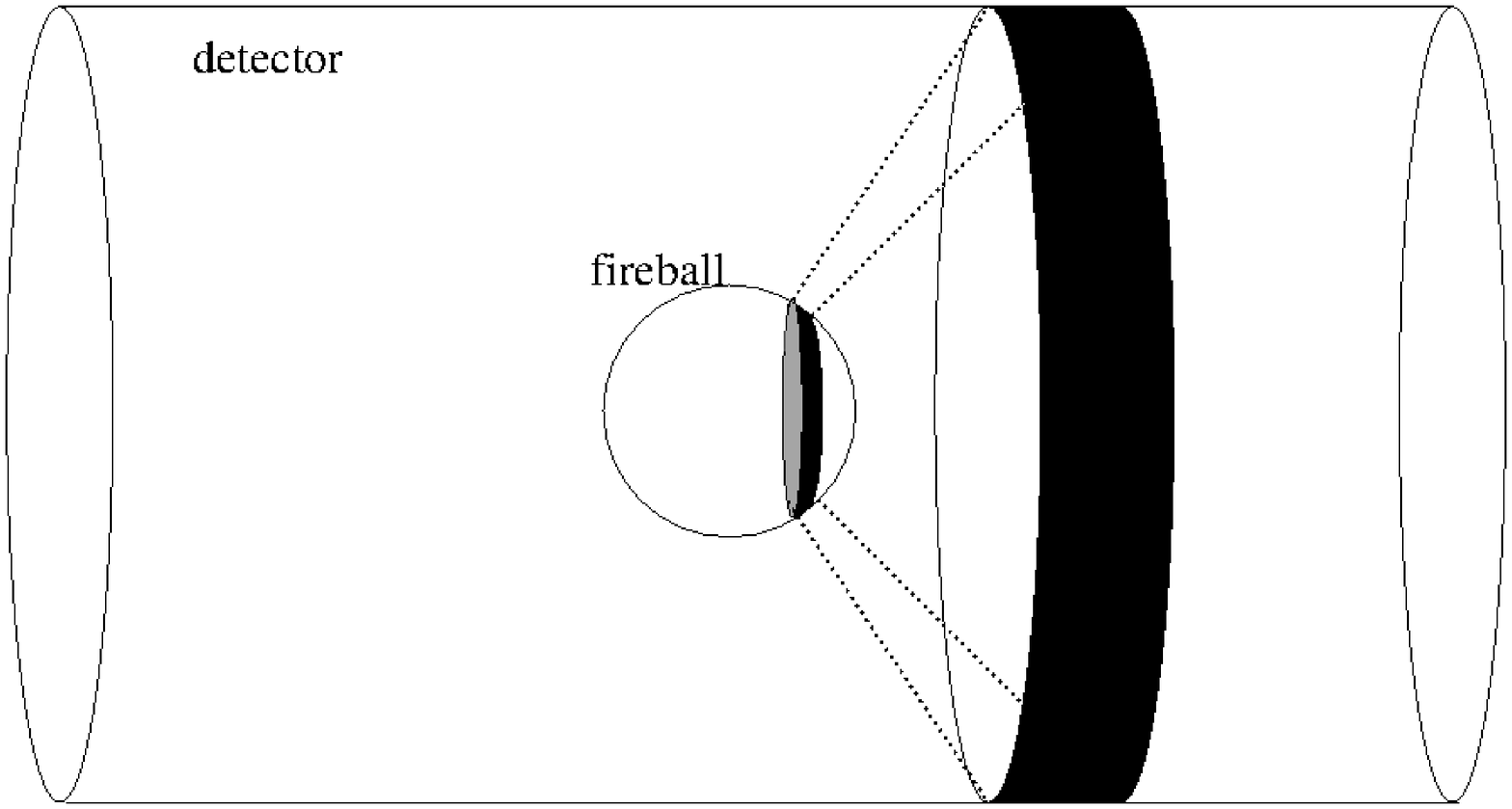}
\end{center}
\caption{Selecting a volume in the fireball by selecting detector cuts
 works best when the momentum of particles in the volume are not totally
 arbitrary. Since hydrodynamics works, we know that in a small volume
 these momenta are aligned with the local fluid momentum, being smeared
 only by an amount of order $T$.}
\label{fg.ensemble}\eef

The least restrictive ensemble for the study of bulk matter is the
microcanonical ensemble. All that this requires is that the energy of
a system be fixed. In fact, since the fireball is well-separated from
the spectators, one may expect a mapping between the collisions and
microcanonical ensemble to be good. However there are two obstructions,
neither of them absolute. The first is that the microcanonical ensemble
requires the energy of each member of the ensemble to be the same.
This is hard to ensure without more control on centrality fluctuations
than is possible at present. Secondly, one needs $4\pi$-detectors to
capture the entire energy of the fireball. Most detectors in use today
miss a very large fraction of the energy.

So one must to try to map the ensemble of events recorded in the
detector to either the canonical or a {\sl grand-canonical ensemble}. The
difference between these is that the system being studied must exchange
either energy or material and energy (respectively) with a much larger
system called a heat-bath. Since detectors accept particles from only
a part of each fireball, one may be able to map the events on to a
grand-canonical ensemble. Of course, thermal and chemical equilibrium
is necessary in order to be able to do this.

We have already discussed the evidence that there is a degree of thermal
and chemical equilibration at freeze out. So, if one observes a small
part of the fireball, it may be possible to treat it in a grand-canonical
ensemble where the rest of the fireball acts as the heat-bath.  In order
to make sure that the system (observed fraction of the fireball) is much
smaller than the heat-bath (the unobserved fraction), one should use as
small an angular coverage as possible while keeping the observed volume
much larger than any intrinsic correlation volumes in the fireball (see
Figure \ref{fg.ensemble}). If the acceptance in rapidity is $\Delta
y$, $V_s$ is the size of the system, and $V_b$ is the volume of the
heat-bath, then
\beq
   \frac{V_s}{V_s+V_b} = \frac{\Delta y}{2\log(\sqrt S/M_p)}\,.
\label{obsvol}\eeq
Taking $\Delta y=2$, one sees that $V_b/V_s$ is about 4.3 at the top
RHIC energy of 200 GeV, and around 7.5 at $\sqrt S=5$ TeV. These may
be acceptable numbers. However, at $\sqrt S=20$ GeV the ratio drops
to 2, and by $\sqrt S=5$ GeV, the ``heat-bath'' is smaller than the
``system''. In order to keep the ratio $V_b/V_s$ fixed, one has to
decrease $\Delta y$ with the beam energy.

This may give rise to another problem, which is to keep the observed
volume much larger than correlation lengths. If freezeout occurs at
time $\tau_f$, then the acceptance region, $\Delta y$, corresponds
approximately to a distance $\Delta x=\tau_f\sinh(\Delta y)$. As long
as correlation lengths are linear in the inverse freezeout temperature
$1/T_f$, it is interesting to examine
\beq
  \Delta x T_f = (\tau_f T_f) \sinh\left(\frac{2\log\sqrt S}{1+V_b/V_s}
    \right).
\label{dimless}\eeq
If one wants $V_b/V_s\simeq4$ at $\sqrt S=5$, where $T_f\simeq145$
MeV, and one takes $\tau_f\simeq5$ fm, then one finds $\Delta x
T_f\simeq2.5$. This is a reasonable number, but it implies that
$\Delta y=0.65$ at this energy.  Such a small acceptance window may
cause statistics to drop significantly.  However, for $\sqrt S\ge20$ GeV,
there is a good possibility that all these constraints may be satisfied
simultaneously.  Of course, if correlation lengths become very large at
some $\sqrt S$ then all these arguments fail, and the system cannot be
treated as being in equilibrium.

Conserved quantities, such as the net particle number or energy,
can change by transport across the boundary of the system.  As a
result, energy and net particle numbers fluctuate in grand-canonical
thermodynamics.  These fluctuations can now be mapped into 
E/E fluctuations. They were first discussed and suggested as
probes of the phase structure of QCD in \cite{qcdebe}. The experimental
variables which allow a direct comparison of QCD predictions with
data were first discovered in \cite{m0123}, and the first lattice QCD
predictions were made in \cite{ilgtiebe}.

\bef[t]
\begin{center}
\includegraphics[scale=0.4]{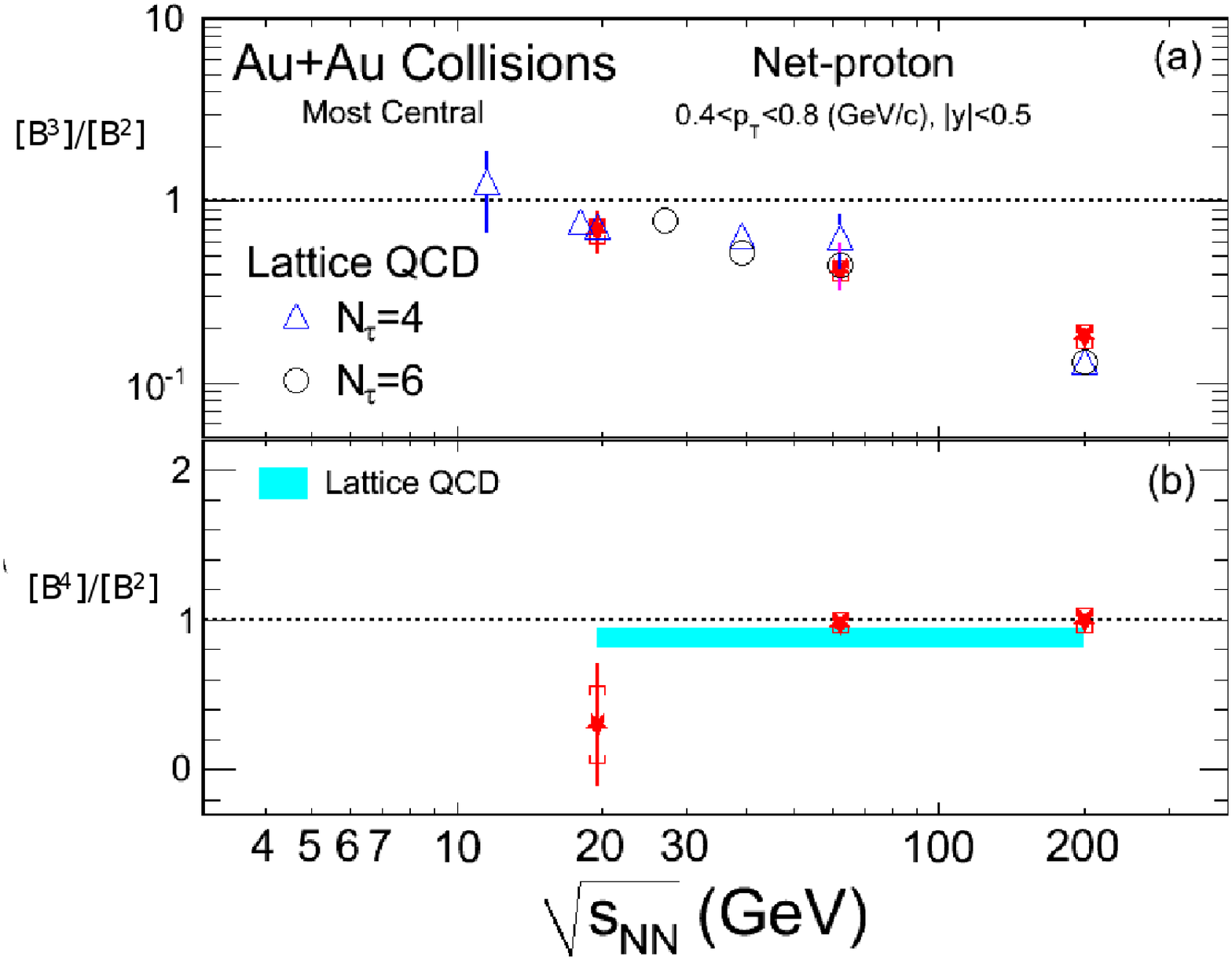}
\end{center}
\caption{A comparison of data on fluctuations of net proton number in the
 STAR experiment with the lattice QCD computations of \cite{ilgtiebe}
 from \cite{starauau}. }
\label{fg.starebe}\eef

The existence of fluctuations means that the baryon number or energy of
an ensemble is not a fixed quantity, but has a probability distribution.
Such a distribution is characterized by the cumulants, $[B^n]$, which
are defined as the Taylor coefficients of the logarithm of the Laplace
transform of the distribution, $P(B)$, of the baryon number---
\beq
   \log\left[\int\,dB\,P(B)\,{\rm e}^{-sB}\right] =
      \sum_{n=1}^\infty [B^n] \frac{(-s)^n}{n!}
\label{cumu}\eeq
The cumulants are related to the Taylor coefficients of the expansion of the
free energy \cite{m0123} in terms of $\mu_B$ simply as
\beq
   [B^n] = V_s \chi^{(n)}(T_f,\mu_B^f)\,T_f^{n-1},
\label{m0123}\eeq
where $\chi^{(n)}(T,\mu_B)$ are generalized quark number susceptibilities
($\chi^{(1)}$ is the baryon density) \cite{taylor}.  As a result, ratios
of the cumulants are independent of the factor $V_s$.  These ratios depend
on the ratios of the dimensionless quantities $\chi^{(n)}(T_f,\mu_B^f)
T_f^{n-4}$, which can be computed in lattice QCD, as demonstrated in
\cite{ilgtiebe}. Similar ratios were also discussed in \cite{atha}.

Since $T_f$ and $\mu_B^f$ was already known from the analysis
of yields, the experimental data could be compared to the lattice
computation \cite{starauau}. This comparison is reproduced in Figure
\ref{fg.starebe}. The remarkably good agreement has led to subsequent
attempts to refine the comparison. These include following up the
suggestions in \cite{ilgtiebe} that the comparison could yield a
measurement of $T_c$ given $T_f$ and $\mu_B^f$ \cite{glmrx} or the
determination of $T_f$ and $\mu_B^f$ given $T_c$ \cite{latticefo}. There
has also been a lot of work on various corrections which may need to
be applied to the experimental data before comparing with predictions
\cite{corr}.  There is also sustained interest in fluid dynamical effects
\cite{Bhalerao0901,fluid}.  In the meanwhile, much new experimental data
has been added from the RHIC Beam Energy Scan (BES) \cite{expt}. A BES-II
is expected shortly.

The long-term goal of the study of fluctuations is to understand
the evolution of fluctuations along the freeze-out curve traced by
changing the beam energy \cite{m0123}. At higher energies one sees
preliminary agreement between lattice predictions and experimental
observations. This agreement is expected to break down in the vicinity
of the QCD critical point because correlation lengths and relaxation
times grow \cite{berdnikov}. At lower energies one might expect a return
to roughly thermal behaviour, although extracting this is fraught with
theoretical and experimental challenges. The BES program aims to locate
and study bulk matter near the QCD critical point.

\section{Acknowledgement}

I would like to thank the organizers of AEPSHEP 2014 for building a very
stimulating scientific program, and for the wonderful local organization.
It was a pleasure to lecture in this school.  I would like to thank the
Kavli Institute for Theoretical Physics China for hospitality during a
part of the time that this manuscript was in preparation. I would like
to thank Bedangadas Mohanty and Rishi Sharma for their helpful comments
on the manuscript.


\begin{thebibliography}{99}
\bibitem{weinberg}
 S.\ Weinberg, {\sl Quantum Theory of Fields: Modern Applications\/}, Vol II,
 Cambridge University Press (2010) Cambridge, Great Britain.
\bibitem{crossover}
 R.\ D.\ Pisarski and F.\ Wilczek, {\sl Phys.\ Rev.\/} D 29 (1984) 338;
 Y.\ Aoki \etal, {\sl Phys.\ Lett.\/} B 643 (2006) 46 [hep-lat/0609068];
 A.\ Bazavov \etal, {\sl Phys.\ Rev.\/} D 85 (2012) 054503 [arxiv:1111.1710].
\bibitem{phased}
 J.\ Berges and K.\ Rajagopal, {\sl Nucl.\ Phys.\/} B 538 (1999) 215 [hep-ph/9804233];
 A.\ M.\ Halasz \etal, {\sl Phys.\ Rev.\/} D 58 (1998) 096007 [hep-ph/9804290].
\bibitem{isospin}
 D.\ T.\ Son and M.\ A.\ Stephanov, {\sl Phys.\ Rev.\ Lett.\/} 86 (2001) 592 [hep-ph/0005225].
\bibitem{sgupta}
 R.\ V.\ Gavai and S.\ Gupta, {\sl Phys.\ Rev.\/} D 66 (2002) 094510 [hep-lat/0208019];
 S.\ Gupta, arxiv:0712.0434.
\bibitem{landau}
 L.\ D.\ Landau and E.\ M.\ Lifshitz, {\sl Course of Theoretical Physics:
  Statistical Physics\/}, Vol 5, Butterworth-Heinemann (1980) Oxford,
  Great Britain.
\bibitem{lattice}
 Ph.\ de Forcrand, {\sl PoS\/} LATTICE2009 (2009) 010 [arxiv:1005.0539];
 S.\ Gupta, {\sl PoS\/} LATTICE2010 (2010) 007 [arxiv:1101.0109];
 L.\ Levkova, {\sl PoS\/} LATTICE2011 (2011) 011 [arxiv:1201.1516];
 G.\ Aarts, {\sl PoS\/} LATTICE 2012 (2012) 017 [arxiv:1302.3028];
 C.\ Gattringer, {\sl PoS\/} LATTICE2013 (2014) 002 [arxiv:1401.7788].
\bibitem{ilgti}
 R.\ V.\ Gavai and S.\ Gupta, {\sl Phys.\ Rev.\/} D 71 (2005) 114014 [hep-lat/0412035];
 R.\ V.\ Gavai and S.\ Gupta, {\sl Phys.\ Rev.\/} D 78 (2008) 114503 [arxiv:0806.2233];
 S.\ Datta \etal, {\sl PoS\/} LATTICE2013 (2014) 202;
 S.\ Gupta \etal, {\sl Phys.\ Rev.\/} D 90 (2014) 034001 [arxiv:1405.2206].
\bibitem{axial}
 R.\ V.\ Gavai \etal, {\sl Phys.\ Rev,\/} D 77 (2008) 114506 [arxiv:0803.0182];
 A.\ Bazavov \etal, {\sl Phys.\ Rev,\/} D 86 (2012) 094503 [arxiv:1205.3535];
 C.\ Bonati \etal, {\sl Phys.\ Rev.\ Lett.\/} 110 (2013) 252003 [arxiv:1301.7640];
 G.\ Cossu \etal, {\sl Phys.\ Rev.\/} D 87 (2013) 114514 [arxiv:1304.6145];
 M.\ I.\ Buchoff \etal, {\sl Phys.\ Rev.\/} D 89 (2014) 054514 [arxiv:1309.4149];
 T.-W.\ Chiu \etal, {\sl PoS\/} LATTICE2013 (2014) 165 [arxiv:1311.6220];
 V.\ Dick \etal, {\sl Phys.\ Rev.\/} D 91 (2015) 094504 [arxiv:1502.06190].
\bibitem{magfield}
 M.\ D'Elia \etal, {\sl Phys.\ Rev.\/} D 82 (2010) 051501 [arxiv:1005.5365];
 G.\ S.\ Bali \etal, {\sl J.\ H.\ E.\ P.\/} 1202 (2012) 044 [arxiv:1111.4956];
 V.\ Skokov, {\sl Phys.\ Rev.\/} D 85 (2012) 034026 [arxiv:1112.5137];
 E.-M.\ Ilgenfritz \etal {\sl Phys.\ Rev.\/} D 85 (2012) 114504 [arxiv:1203.3360];
 F.\ Bruckmann \etal, {\sl J.\ H.\ E.\ P.\/} 1304 (2013) 112 [arxiv:1303.3972];
 L.\ Levkova and C.\ De Tar, {\sl Phys.\ Rev.\ Lett.\/} D 112 (2014) 012002 [arxiv:1309.1142];
 C.\ Bonati \etal {\sl Phys.\ Rev.\/} D 89 (2014) 054506 [arxiv:1310.8656];
 G.\ S.\ Bali \etal, {\sl J.\ H.\ E.\ P.\/} 1408 (2014) 177 [arxiv:1406.0269];
\bibitem{thermalize}
 R.\ Baier \etal, {\sl Phys.\ Lett.\/} B 502 (2001) 51 [hep-ph/0009237];
 A.\ Dumitru \etal, {\sl Phys. Rev.\/} D 75 (2007) 025016 [hep-ph/0604149];
 P.\ Romatschke and R.\ Venugopalan, {\sl Phys.\ Rev.\/} D 74 (2006) 045011;
 Y.\ V.\ Kovchegov and A.\ Taliotis, {\sl Phys.\ Rev.\/} C 76 (2007) 014905 [arxiv:0705.1234];
 D.\ A.\ Teaney, arxiv:0905.2433;
 V.\ Balasubramanian \etal, {\sl Phys.\ Rev.\/} 84 (2011) 026010 [arxiv:1103.2683];
 A.\ Kurkela and G.\ D.\ Moore, {\sl J.\ H.\ E.\ P.\/} 1112 (2011) 044 [arxiv:1107:5050];
 K.\ Fukushima and F.\ Gelis, {\sl Nucl.\ Phys.\/} A 874 (2012) 108 [arxiv:1106.1396];
 T.\ Epelbaum and F.\ Gelis, {\sl Nucl.\ Phys.\/} A 872 (2011) 210 [arxiv:1107.0668];
 J.-P.\ Blaizot \etal, {\sl Nucl.\ Phys.\/} A 873 (2012) 68 [arxiv: 1107.5296];
 J.\ Berges \etal, {\sl Phys.\ Rev.\/} D 89 (2014) 7, 074011 [arxiv: 1303:5650].
\bibitem{general}
 N.\ Armesto \etal, {\sl J.\ Phys.\/} G 35 (2008) 054001 [arxiv:0711.0974];
 B.\ Miller \etal, {\sl Ann.\ Rev.\ Nucl.\ Part.\ Sci.\/} 62 (2012) 361 [arxiv:1202.3233].
\bibitem{Gale1209}
 C.\ Gale \etal, {\sl Phys.\ Rev.\ Lett.\/} 110 (2013) 012302 [arxiv:1209.6330].
\bibitem{ALICE1011}
 K.\ Aamodt \etal, {\sl Phys.\ Rev.\ Lett.\/} 105 (2010) 252301 [arxiv:1011.3916].
\bibitem{jetq}
 M.\ Gyulassy and M.\ Pl\"umer, {\sl Phys.\ Lett.\/} B 243 (1990) 432;
 X.-N.\ Wang and M.\ Gyulassy, {\sl Phys.\ Rev.\ Lett.\/} 68 (1992) 1480.
\bibitem{STAR0306}
 STAR Collaboration, {\sl Phys.\ Rev.\ Lett.\/} 91 (2003) 072304 [arxiv:nucl-ex/0306024]
\bibitem{CMS1102}
 S.\ Chatrchyan \etal (CMS Collaboration), {\sl Phys.\ Rev.\ Lett.\/} 106 (2011) 212301 [arxiv:1102.5435].
\bibitem{Lee2014}
 Yan-Jie Lee, QM 2014: https://indico.cern.ch/ event/ 219436/ session/ 3/ contribution/ 728/ material/ slides/ 1.pdf
\bibitem{Baier}
 R.\ Baier \etal, {\sl Phys.\ Lett.\/} B 345 (1995) 277 [hep-ph/9411409];
 P.\ B.\ Arnold \etal, {\sl J.\ H.\ E.\ P.\/} 0206 (2002) 030 [hep-ph/0204343];
 G.\ Baym \etal, {\sl Phys.\ Lett.\/} B 644 (2007) 48 [hep-ph/0604209];
 S.\ Peigne and A.\ V.\ Smilga, {\sl Phys.\ Usp.\/} 52 (2009) 659 [arxiv:0810.5702];
 S.\ Caron-Huot, {\sl Phys.\ Rev.\/} D 79 (2009) 065039 [arxiv:0811.1603].
 J.\ Ghiglieri and D.\ Teaney, arxiv:1502.03730.
\bibitem{Majumder}
 A.\ Majumder, {\sl Phys.\ Rev.\/} C 87 (2013) 034905 [arxiv:1202.5295];
 X.\ Ji, {\sl Phys.\ Rev.\ Lett.\/} 110 (2013) 262002 [arxiv:1305.1539];
 M.\ Panero \etal, {\sl Phys.\ Rev.\ Lett.\/} 112 (2014) 162001 [arxiv:1307.5850];
 M.\ Laine and A.\ Rothkopf, {\sl PoS\/} LATTICE2013 (2013) 174 [arxiv:1310.2413].
\bibitem{Wiedemann}
 H.\ Liu \etal, {\sl Phys.\ Rev.\ Lett.\/} 98 (2007) 182301 [hep-ph/0607062];
 H.\ Liu \etal, {\sl J.\ H.\ E.\ P.\/} 0703 (2007) 066 [hep-ph/0612168];
 J.\ Casalderrey-Solana and D.\ Teaney, {\sl J.\ H.\ E.\ P.\/} 0704 (2007) 039 [hep-th/0701123];
 Y.\ Hatta, E.\ Iancu and A.\ H.\ Mueller {\sl J.\ H.\ E.\ P.\/} 0805 (2008) 037 [arxiv:0803.2481];
 P.\ M.\ Chesler \etal, {\sl Phys.\ Rev.\/} D 79 (2009) 125015 [arxiv:0810.1985];
 J.\ de Boer \etal, {\sl J.\ H.\ E.\ P.\/} 0907 (2009) 094 [arxiv:0812.5112].
\bibitem{bj}
 J.\ D.\ Bjorken, {\sl Phys.\ Rev.\/} D 27 (1983) 140.
\bibitem{Ollitrault}
 J.-Y.\ Ollitrault, {\sl Phys.\ Rev.\/} D 46 (1992) 229.
\bibitem{Voloshin}
 S.\ Voloshin and Y.\ Zhang, {\sl Z.\ Phys.\/} C 70 (1996) 665 [hep-ph/9407282].
\bibitem{Srivastava}
 A.\ P.\ Mishra \etal, {\sl Phys.\ Rev.\/} C 81 (2010) 034903 [arxiv:0811.0292];
 B.\ Alver and G.\ Roland, {\sl Phys.\ Rev.\/} C 81 (2010) 054905 [arxiv:1003.0194];
 D.\ Teaney and Li Yan, {\sl Phys.\ Rev.\/} C 83 (2011) 064904 [arxiv:1010.1876].
\bibitem{Dumitru}
 H.\-J.\ Dresher \etal, {\sl Phys.\ Rev./} C 76 (2007) 024905 [arxiv:0704.3553];
 P.\ Romatschke and U.\ Romatschke, {\sl Phys.\ Rev.\ Lett.\/} 99  (2007) 172301 [arxiv:0706.1522];
 H.\ Song and U.\ W.\ Heinz, {\sl Phys.\ Lett.\/} B 658 (2008) 279 [arxiv:0709.0742].
 K.\ Dusling and D.\ Teaney, {\sl Phys.\ Rev.\/} C 77 (2008) 034905 [arxiv:0710.5932]
\bibitem{flow}
 M.\ Luzum and P.\ Romatschke, {\sl Phys.\ Rev.\/} C 78 (2008) 034915 [arxiv:0804.4015];
 G.\ Ferini \etal, {\sl Phys.\ Lett.\/} B 670 (2009) 325 [arxiv:0805.4814];
 T.\ Hirano and Y.\ Nara, {\sl Phys.\ Rev.\/} C 79 (2009) 064904 [arxiv:0904.4080];
 K.\ Dusling \etal, {\sl Phys.\ Rev.\/} C 81 (2010) 034907 [arxiv:0909.0754];
 B.\ Schenke \etal, {\sl Phys.\ Rev.\/} C 82 (2010) 014903 [arxiv:1004.1408];
 H.\ Holopainen \etal, {\sl Phys.\ Rev.\/} C 83 (2011) 034901 [arxiv:1007.0368];
 G.\-Y.\ Qin \etal, {\sl Phys.\ Rev.\/} C 82 (2010) 064903 [arxiv:1009.1847];
 C.\ Shen \etal, {\sl Phys.\ Rev.\/} C 84 (2011) 044903 [arxiv:1105.3226];
 P.\ Bozek, {\sl Phys.\ Rev.\/} C 85 (2012) 034901 [arxiv:1110.6742];
 M.\ Martinez \etal, {\sl Phys.\ Rev.\/} C 85 (2012) 064913 [arxiv:1204.1473];
 D.\ Teaney and Li Yan, {\sl Phys.\ Rev.\/} C 86 (2012) 044908 [arxiv:1206.1905];
 STAR Collaboration, {\sl Phys.\ Rev.\/} C 88 (2013) 014904 [arxiv:1301.2187];
 CMS Collaboration, {\sl Phys.\ Lett.\/} B 724 (2013) 213 [arxiv:1305.0609];
 ATLAS Collaboration, {\sl J.\ H.\ E.\ P.\/} 1311 (2013) 183 [arxiv:1305.2942];
 ALICE Collaboration, {\sl Phys.\ Lett.\/} B 726 (2013) 164 [arxiv:1307.3237].
\bibitem{Cleymans}
 J.\ Cleymans and K.\ Redlich, {\sl Phys.\ Rev.\ Lett.\/} 81 (1998) 5284 [nucl-th/9808030];
 Z.-W.\ Lin \etal, {\sl Phys.\ Rev.\/} C 72 (2005) 064901 [nucl-th/064901];
 A.\ Andronic \etal, {\sl Nucl.\ Phys.\/} A 772 (2006) 167 [nucl-th/0511071];
 F.\ Becattini \etal, {\sl Phys.\ Rev.\/} C73 (2006) 044905 [hep-ph/0511092];
 NA49 Collaboration {\sl Phys.\ Rev.\/} C 73 (2006) 044910;
 V.\ V.\ Begun \etal, {\sl Phys.\ Rev.\/} C 76 (2007) 024902 [nucl-th/0611075];
 F.\ Becattini and J.\ Manninen, {\sl J.\ Phys.\/} G 35 (2008) 104013 [arxiv:0805.0098];
 STAR Collaboration, arxiv:1007.2613;
 M.\ Chojnacki \etal, {\sl Comput.\ Phys.\ Commun.\/} 183 (2012) 746 [arxiv:1102.0273];
 ALICE Collaboration, {\sl Phys.\ Rev.\/} C 88 (2013) 044910 [arxiv:1303.0737];
 S.\ Borsanyi \etal, {\sl Phys.\ Rev.\ Lett.\/} 111 (2013) 062005 [arxiv:1305.5161].
\bibitem{Bhalerao0901}
 R.\ S.\ Bhalerao and S.\ Gupta, {\sl Phys.\ Rev.\/} C 79 (2009) 064901 [arxiv:0901.4677].
\bibitem{2cfo}
 S.\ Chatterjee \etal, {\sl Phys.\ Lett.\/} B 727 (2013) 554 [arxiv:1306.2006];
 K.\ A.\ Bugaev \etal, {\sl Europhys.\ Lett.\/} 104 (2013) 22002 [arxiv:1308.3594].
\bibitem{nonthermal}
 J.\ Steinheimer \etal, {\sl Phys.\ Rev.\ Lett.\/} 110 (2013) 042501 [arxiv:1203.5302];
 F.\ Becattini \etal {\sl Phys.\ Rev.\ Lett.\/} 111 (2013) 082302 [arxiv:1212.2431].
\bibitem{matsui}
 T.\ Matsui and H.\ Satz, {\sl Phys.\ Lett.\/} B 178 (1986) 416.
\bibitem{jpsilat}
 M.\ Asakawa and T.\ Hatsuda, {\sl Phys.\ Rev.\ Lett.\/} 92 (2004) 012001 [hep-lat/0308034];
 S.\ Datta \etal, {\sl Phys.\ Rev.\/} D 69 (2004) 094507 [hep-lat/0312037].
\bibitem{stepwise}
 F.\ Karsch \etal, {\sl Phys.\ Lett.\/} B 637 (2006) 75 [hep-ph/0512239];
 H.\ Satz, {\sl Nucl.\ Phys.\/} A 783 (2007) 249.
\bibitem{upsdat}
 S.\ Chatrchyan \etal {\sl Phys.\ Rev.\ Lett.\/} 109 (2012) 222301 [arxiv:1208.2826];
 S.\ Chatrchyan \etal {\sl Phys.\ Rev.\ Lett.\/} 107 (2011) 052302 [arxiv:1105.4894].
\bibitem{Sharma1401}
 S.\ Gupta and R.\ Sharma, {\sl Phys.\ Rev.\/} C 89 (2014) 057901 [arxiv:1401.2930].
\bibitem{jpsidyn}
 N.\ Brambilla \etal, {\sl Eur.\ Phys.\ J.\/} C 71 (2011) 1534, and references
 therein.
\bibitem{qcdebe}
 M.\ A.\ Stephanov \etal, {\sl Phys.\ Rev.\ Lett.\/} 81 (1998) 4816 [hep-ph/9806219];
 M.\ A.\ Stephanov \etal, {\sl Phys.\ Rev.\/} D 60 (1999) 114028 [hep-ph/9903292].
\bibitem{m0123}
 S.\ Gupta, {\sl PoS\/} CPOD2009 (2009) 025 [arxiv:0909.4630];
 S.\ Gupta, {\sl Prog.\ Theor.\ Phys.\ Suppl.\/} 186 (2010) 440.
\bibitem{ilgtiebe}
 R.\ V.\ Gavai and S.\ Gupta, {\sl Phys.\ Lett.\/} B 696 (2011) 459 [arxiv:1001.3796].
\bibitem{taylor}
 R.\ V.\ Gavai and S.\ Gupta, {\sl Phys.\ Rev.\/} D 68 (2003) 034506 [hep-lat/0303013].
\bibitem{atha}
 C.\ Athanasiou \etal, {\sl Phys.\ Rev.\/} D 82 (2010) 074008 [arxiv:1006.4636].
\bibitem{starauau}
 M.\ M.\ Aggarwal \etal (STAR Collaboration), {\sl Phys.\ Rev.\ Lett.\/} 105 (2010) 022302 [arxiv:1004.4959].
\bibitem{glmrx}
 S.\ Gupta \etal, {\sl Science\/} 332 (2011) 1525 [arxiv:1105.3934].
\bibitem{latticefo}
 A.\ Bazavov \etal, {\sl Phys.\ Rev.\ Lett.\/} 109 (2012) 192302 [arxiv:1208.1220];
 S.\ Borsanyi \etal, {\sl Phys.\ Rev.\ Lett.\/} 111 (2013) 062005 [arxiv:1305.5161].
\bibitem{corr}
 E.\ S.\ Fraga \etal, {\sl Phys.\ Rev.\/} C 84 (2011) 011903 [arxiv:1104.3755];
 M.\ Kitazawa and M.\ Asakawa {\sl Phys.\ Rev.\/} C 85 (2012) 021901 [arxiv:1107.1412];
 A.\ Bzdak etal, {\sl Phys.\ Rev.\/} C 87 (2013) 014901 [arxiv:1203.4529];
 M.\ Kitazawa and M.\ Asakawa {\sl Phys.\ Rev.\/} C 86 (2012) 024904 [arxiv:1205.3292];
 V.\ Skokov and B.\ Friman, {\sl Phys.\ Rev.\/} C 88 (2013) 034911 [arxiv:1205.4756];
 A.\ Bzdak and V.\ Koch, {\sl Phys.\ Rev.\/} C 86 (2012) 044904 [arxiv:1206.4286];
 X.\ Luo \etal, {\sl J.\ Phys.\/} G 40 (2013) 105104 [arxiv:1302.2332];
 H.\ Ono \etal, {\sl Phys.\ Rev.\/} C 87 (2013) 041901 [arxiv:1303.3338];
 P.\ Garg \etal, {\sl Phys.\ Lett.\/} B 726 (2013) 691 [arxiv:1304.7133];
 A.\ Tang and G.\ Wang, {\sl Phys.\ Rev.\/} C 88 (2013) 024905 [arxiv:1305.1392];
 L.\ Chen \etal, {\sl J.\ Phys.\/} G 41 (2014) 105107 [arxiv:1312.0749];
 A.\ Bzdak and V.\ Koch, {\sl Phys.\ Rev.\/} C 91 (2015) 027901 [arxiv:1312.4574];
 X.\ Luo \etal, {\sl Nucl.\ Phys.\/} A 931 (2014) 808 [arxiv:1408.0495];
 M.\ Sakaida \etal, {\sl Phys.\ Rev.\/} C 90 2014) 064911 [arxiv:1409.6866].
\bibitem{fluid}
 M.\ A.\ Stephanov, {\sl Phys.\ Rev.\/} D 81 (2010) 054012 [arxiv:0911.1772];
 K.\ Xiao \etal, {\sl Chin.\ Phys.\/} C 35 (2011) 467;
 J.\ I.\ Kapusta and J.\ M.\ Torres-Rincon, {\sl Phys.\ Rev.\/} C 86 (2012) 054911 [arxiv:1209.0675];
 M.\ Kitazawa, arxiv: 1505.04349;
 S.\ Mukherjee \etal, arxiv:1506.00645
\bibitem{expt}
 X.\ Zhang (STAR Collaboration) {\sl Nucl.\ Phys.\/} A 904-905 (2013) 543c;
 A.\ Sarkar (STAR Collaboration) {\sl PoS\/} CPOD2013 (2013) 043;
 L.\ Adamczyk \etal (STAR Collaboration) {\sl Phys.\ Rev.\ Lett.\/} 112 (2014) 032302 [arxiv:1309.5681];
 L.\ Adamczyk \etal (STAR Collaboration) {\sl Phys.\ Rev.\ Lett.\/} 113 (2014) 092301 [arxiv:1402.1558];
 A.\ Adare \etal (PHENIX Collaboration) arxiv:1506.07834.
\bibitem{berdnikov}
 B.\ Berdnikov and K.\ Rajagopal, {\sl Phys.\ Rev.\/} D 61 (2000) 105017 [hep-ph/9912274];
 M.\ A.\ Stephanov, {\sl Phys.\ Rev.\ Lett.\/} 102 (2009) 032301 [arxiv:0809.3450].
\end{thebibliography}
\end{document}